\documentclass[lettersize,journal]{IEEEtran}
\usepackage[footnotesize,labelfont=bf]{caption}
\usepackage[caption=false,font=normalsize,labelfont=sf,textfont=sf]{subfig}
\def\BibTeX{{\rm B\kern-.05em{\sc i\kern-.025em b}\kern-.08em
    T\kern-.1667em\lower.7ex\hbox{E}\kern-.125emX}}
\usepackage{balance}

\def\plotFlag{true} 
\def\arxivFlag{true} 

\usepackage[table, dvipsnames]{xcolor} 
\usepackage[nogroupskip, nopostdot]{glossaries} 
\usepackage{amsmath,amsfonts}
\usepackage{bm}
\usepackage{graphicx}
\usepackage{xfrac}
\usepackage{siunitx}
\usepackage{layouts}
\usepackage{tabularx}
\usepackage{subcaption}
\usepackage{pgf}
\usepackage{tikz}
\usepackage[export]{adjustbox} 
\usepackage{enumitem} 
\usepackage{tcolorbox} 
\usepackage{xstring} 
\usepackage{pgfplots} 
\usepackage{pgfplotstable} 
\usepackage{booktabs}
\usepackage{multirow}
\usepackage{makecell} 
\usepackage{url}
\usepackage[backend=biber,url=false,doi=false,isbn=false,style=ieee]{biblatex} 
\usepackage[normalem]{ulem} 
\usepackage{xspace} 
\usepackage{lipsum} 
\usepackage{orcidlink} 
\usepackage[capitalise]{cleveref} 

\newacronym{ar}{AR}{augmented reality}
\newacronym{hmd}{HMD}{head-mounted display}
\newacronym{aoi}{AoI}{age of information}
\newacronym{bpc}{BPC}{bits per color}
\newacronym{ifs}{IFS}{inter-frame spacing}
\newacronym{imu}{IMU}{inertial measurement unit}
\newacronym{fps}{FPS}{frames per second}
\newacronym{qoe}{QoE}{quality of experience}
\newacronym{aqoe}{AppQoE}{application \gls{qoe}}
\newacronym{vr}{VR}{virtual reality}
\newacronym{xr}{XR}{extended reality}
\newacronym{vr/ar}{VR/AR}{virtual/augmented reality}
\newacronym{6dof}{6DoF}{six degrees of freedom}

\newacronym{ip}{IP}{internet protocol}
\newacronym{lan}{LAN}{local area network}
\newacronym{m2p}{M2P}{motion-to-photon}
\newacronym{qos}{QoS}{quality of service}
\newacronym{rtp}{RTP}{real-time transport protocol}
\newacronym{udp}{UDP}{user datagram protocol}
\newacronym{wifi}{Wi-Fi}{The Standard for Wireless Fidelity (IEEE 802.11 family)}

\newacronym{ack}{ACK}{acknowledgement}
\newacronym{ampdu}{A-MPDU}{aggregated MAC protocol data unit}
\newacronym{amsdu}{A-MSDU}{aggregated \gls{msdu}}
\newacronym{a-bft}{A-BFT}{association beam forming training}
\newacronym{bft}{BFT}{beamforming training}
\newacronym{bf}{BF}{beamforming}
\newacronym{bi}{BI}{beacon interval}
\newacronym{brp}{BRP}{beam refinement period}
\newacronym{br}{BR}{beam refinement}
\newacronym{bss}{BSS}{basic service set}
\newacronym{bti}{BTI}{beacon transmission interval}
\newacronym{bt}{BT}{beam tracking}
\newacronym{cbap}{CBAP}{contention-based access period}
\newacronym{cts}{CTS}{clear to send}
\newacronym{dti}{DTI}{data transmission interval}
\newacronym{rtxss}{\mbox{R-TXSS}}{responder transmit sector sweep}
\newacronym{sls}{SLS}{sector-level sweep}
\newacronym{sp}{SP}{service period}
\newacronym{ssw}{SSW}{sector sweep}
\newacronym{sta}{STA}{station}
\newacronym{stp}{STP}{setup message}
\newacronym{tu}{TU}{time unit}
\newacronym{txss}{TXSS}{transmit sector sweep}
\newacronym{trn-t}{TRN-T}{training sequence}
\newacronym{trn-r}{TRN-R}{training sequence}
\newacronym{tru}{TRU}{training unit}
\newacronym{thr}{THR}{throughput}
\newacronym{fbk}{FBK}{feedback}

\newacronym{rx}{RX}{receiver}
\newacronym{aoa}{AoA}{angle of arrival}
\newacronym{aod}{AoD}{angle of departure}
\newacronym{eoa}{EoA}{elevation of arrival}
\newacronym{ber}{BER}{bit error ratio}
\newacronym{e2e}{E2E}{end-to-end}
\newacronym{evm}{EVM}{error vector magnitude}
\newacronym{fov}{FOV}{field of view}
\newacronym{gi}{GI}{guard interval}
\newacronym{hpbw}{HPBW}{half power beam width}
\newacronym{ldpc}{LDPC}{low-density parity-check}
\newacronym{los}{LoS}{line of sight}
\newacronym{nlos}{NLoS}{non-line-of-sight}
\newacronym{olos}{OLoS}{obstructed line-of-sight}
\newacronym{mcs}{MCS}{modulation and coding scheme}
\newacronym{mmw}{mmWave}{millimeter-wave}
\newacronym{mpc}{MPC}{multipath component}
\newacronym{phy}{PHY}{physical layer}
\newacronym{ppdu}{PPDU}{PHY protocol data unit}
\newacronym{psdu}{PSDU}{PHY service data unit}
\newacronym{mac}{MAC}{medium access control}
\newacronym{mpdu}{MPDU}{MAC protocol data unit}
\newacronym{msdu}{MSDU}{MAC service data unit}
\newacronym{rat}{RAT}{radio access technology}
\newacronym{rss}{RSS}{received singla strength}
\newacronym{snr}{SNR}{signal-to-noise ratio}
\newacronym{tof}{ToF}{time of flight}
\newacronym{tx}{TX}{transmitter}
\newacronym{upa}{UPA}{uniform planar array}
\newacronym{ura}{URA}{uniform radial array}
\newacronym{eirp}{EIRP}{equivalent isotropically radiated power}
\newacronym{cir}{CIR}{channel impulse response}
\newacronym{ctf}{CTF}{channel transfer function}
\newacronym{mrc}{MRC}{maximum ratio combining}
\newacronym{mrt}{MRT}{maximum ratio transmission}
\newacronym{mrtmrc}{MRT/MRC}{joint maximum ratio transmission/combining}
\newacronym{sm}{SM}{spatial multiplexing}
\newacronym{hh}{HH}{horizontal-to-horizontal}
\newacronym{vv}{VV}{vertical-to-vertical}
\newacronym{agc}{AGC}{automatic gain control}
\newacronym{fspl}{FSPL}{free-space path loss}
\newacronym{ofdm}{OFDM}{orthogonal frequency-division multiplexing}
\newacronym{mimo}{MIMO}{multiple-input multiple-output}
\newacronym{de}{DE}{dominant eigenmode}
\newacronym{fdm}{FDM}{fully-digital \gls{mimo}}
\newacronym{det}{DET}{dominant eigenmode transmission}
\newacronym{dl}{DL}{downlink}
\newacronym{ul}{UL}{uplink}
\newacronym{abg}{ABG}{alpha-beta-gamma}
\newacronym{rf}{RF}{radio frequency}
\newacronym{sage}{SAGE}{space alternating generalized expectation-maximization}
\newacronym{isi}{ISI}{inter-symbol interference}
\newacronym{sc}{SC}{single carrier}
\newacronym{gtd}{GTD}{geometric theory of diffraction}

\newacronym{ap}{AP}{access point}
\newacronym{bs}{BS}{base station}
\newacronym{cots}{COTS}{consumer off-the-shelf}
\newacronym{kpi}{KPI}{key performance indicator}
\newacronym{dmg}{DMG}{directional multi-gigabit}
\newacronym{ue}{UE}{user equipment}
\newacronym{me}{ME}{monitoring entity}
\newacronym{db}{DB}{database}
\newacronym{rl}{RL}{reinforcement learning}
\newacronym{soa}{SoA}{state-of-the-art}
\newacronym{mab}{MAB}{multi-armed bandit}
\newacronym{cdf}{CDF}{cumulative distribution function}
\newacronym{ccdf}{CCDF}{complementary cumulative distribution function}
\newacronym{fft}{FFT}{fast Fourier transform}
\newacronym{gpu}{GPU}{graphics processing unit}
\newacronym{pdf}{PDF}{probability density function}
\newacronym{adbft}{AdBFT}{automatic and dynamic \gls{bft}}
\newacronym{rnd}{R\&D}{research and development}
\newacronym{io}{IO}{input/output}
\newacronym{pmf}{PMF}{probability mass function}
\newacronym{trl}{TRL}{technology readiness level}

\newacronym{sf}{SF}{shadow fading}
\newacronym{hf}{HF}{hand fading}
\newacronym{vf}{VF}{view fading}
\newacronym{mf}{MF}{misalignment fading}

\newacronym{ris}{RIS}{reconfigurable intelligent surface}

\newacronym{ir}{IR}{infrared}

\newacronym{ssq}{SSQ}{simulator sickness questionnaire}

\newacronym{ban}{BAN}{body-area network}

\newacronym{rmse}{RMSE}{root-mean-square error}

\newacronym{slerp}{SLERP}{spherical linear interpolation}

\newacronym{psd}{PSD}{power-spectral density}

\glsdisablehyper 

\crefname{figure}{fig.}{figs.}

\AtBeginBibliography{\footnotesize} 
\AtEveryBibitem{\clearlist{language}}
\AtEveryBibitem{\clearlist{publisher}}
\AtEveryBibitem{\clearlist{location}}

\newcolumntype{Y}{>{\centering\arraybackslash}X}

\usetikzlibrary{calc}
\usetikzlibrary{shapes.geometric}
\usetikzlibrary{shapes.arrows}
\usetikzlibrary{arrows}
\usetikzlibrary{arrows.meta}
\usetikzlibrary{decorations.pathreplacing}
\usetikzlibrary{spy} 
\usepgfplotslibrary{colormaps}
\usetikzlibrary{shapes}
\pgfplotsset{compat=1.18} 
\pgfplotsset{
    colormap={kulmap}{
        rgb255=(0, 64, 122)
        rgb255=(82, 189, 236)
        rgb255=(255, 248, 242)
    }
}

\DeclareSIUnit{\bps}{bps}
\DeclareSIUnit\bit{b}

\definecolor{kullogo1}{RGB}{89, 189, 236}
\definecolor{kullogo2}{RGB}{0, 64, 112}
\definecolor{kulgrid1}{RGB}{31, 171, 231}
\definecolor{kulgrid12}{RGB}{35, 145, 185} 
\definecolor{kulgrid2}{RGB}{29, 141, 176}
\definecolor{kulgrid3}{RGB}{17, 110, 138}

\definecolor{my_red}{RGB}{255, 0, 0}
\definecolor{my_pink}{RGB}{245, 113, 240}

\definecolor{apBlue}{RGB}{0,64,122} 
\definecolor{ueRed}{RGB}{255, 0, 0} 

\definecolor{chpoutlineboxcolor}{RGB}{220, 220, 220} 
\definecolor{chptodoboxcolor}{RGB}{250, 200, 120} 
\definecolor{questionboxcolor}{RGB}{230, 255, 170} 

\definecolor{green1}{RGB}{20,150,20}

\definecolor{hl_gray}{RGB}{160,160,160}
\definecolor{wr_gray}{RGB}{200,200,200}
\definecolor{pw_gray}{RGB}{240,240,240}

\newcommand{\hali}[0]{Half\nobreakdash-Life:~Alyx\xspace}
\newcommand{\wren}[0]{Wrench\xspace}
\newcommand{\piwi}[0]{Pistol Whip\xspace}

\newcommand{\glsposs}[1]{%
  \ifglsused{#1}{%
    \glsdisp{#1}{\glsentryshort{#1}'s}%
  }{%
    \glsdisp{#1}{\glsentrylong{#1}'s (\glsentryshort{#1}'s)}%
  }%
}

\usepackage{xspace}

\def\arxivColor{WildStrawberry}

\newcommand{\arxiv}[1]{%
\IfEqCase{\arxivFlag}{%
    {false}{}%
    {true}{#1}%
    {colored}{{\color{\arxivColor}#1}}%
}[\PackageError{tree}{Undefined option to tree: \arxivflag}{}]%
}

\newcommand{\narxiv}[1]{%
\IfEqCase{\arxivFlag}{%
    {false}{#1}%
    {true}{}%
    {colored}{}%
}[\PackageError{tree}{Undefined option to tree: \arxivflag}{}]%
}

\def\centerarc[#1](#2)(#3:#4:#5){ \draw[#1] ($(#2)+({#5*cos(#3)},{#5*sin(#3)})$) arc (#3:#4:#5); }

\tikzset{
    partial ellipse/.style args={#1:#2:#3}{
        insert path={+ (#1:#3) arc (#1:#2:#3)}
    }
}

\newcommand{\altFigPath}{alt_fig.png}

\newcommand{\plotFig}[2]{
\IfEqCase{#1}{%
    {true}{#2}%
    {false}{\includegraphics[width=\linewidth]{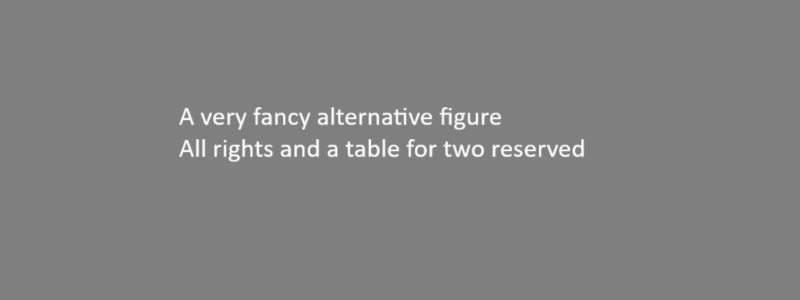}}%
}[\PackageError{tree}{Undefined option to tree: #1}{}]%
}

\newcommand{\gilles}[2][]{%
\IfStrEqCase{#1}{%
    {}{{\color{Plum}[GILLES: #2]}}%
    {left}{{\color{Plum} $\leftarrow$[GILLES: #2]}}%
    {right}{{\color{Plum}[GILLES: #2] $\rightarrow$}}%
}[{Error: Invalid input: #1}]
}

\arxiv{
\usepackage[firstpage=true]{background}
\usepackage{eso-pic}

\backgroundsetup{
  position=current page.south,
  angle=0,
  nodeanchor=south,
  vshift=-0.5mm,
  hshift=-2.6mm,
  opacity=1,
  scale=3,
  color=blue!80!white,
  contents={\begin{varwidth}{\textwidth}\fontsize{3}{3.4}\selectfont 
  © 2024 IEEE. Personal use of this material is permitted. Permission from IEEE must be obtained for all other uses, in any current \\
  or future media, including reprinting/republishing this material for advertising or promotional purposes, creating new collective \\
  works, for resale or redistribution to servers or lists, or reuse of any copyrighted component of this work in other works. \\
  This is the accepted version of the published paper with DOI: {\fontsize{3.4}{14}\selectfont 10.1109/MCOM.001.2300749}\\
  \end{varwidth}
  }
}
}

\pgfplotsset{
  layers/axis lines on top/.define layer set={
    axis background,
    axis grid,
    pre main,
    main,
    axis lines,
    axis descriptions,
    axis ticks,
    axis tick labels,
    axis foreground,
  }{/pgfplots/layers/standard},
  generalPlotStyle/.style={
        width=8.85cm,
        height=6cm,
        every axis/.append style={
            font=\footnotesize\sffamily
        },
        title style={font=\sffamily\normalsize},
        legend style={
            at={(0.02,0.98)},
            anchor=north west,
            font=\footnotesize\sffamily
        },
        tick label style={
            font=\footnotesize\sffamily,
            /pgf/number format/assume math mode=true,
        },
        tick style={black, solid, line width=0.25pt},
        major tick length=2pt,
        legend cell align={left},
        grid,
        grid style={loosely dotted, black, line width=0.5pt},
        extra x tick style={major grid style={draw=none}}
        xlabel style={
            yshift=0pt,
            font=\normalsize\sffamily,
        },
        ylabel style={
            yshift=-4pt,
            font=\normalsize\sffamily,
        },
  }
}

\begin{document}

\pgfplotstableread[col sep=comma]{figure/dir-com/head_position_pdf_x.csv}\headPositionPdfX
\pgfplotstableread[col sep=comma]{figure/dir-com/head_position_pdf_y.csv}\headPositionPdfY
\pgfplotstableread[col sep=comma]{figure/dir-com/head_vertical_factor_pdf.csv}\headVerticalFactorPdf
\pgfplotstableread[col sep=comma]{figure/dir-com/head_orientation_pdf_yaw.csv}\headOrientationPdfYaw
\pgfplotstableread[col sep=comma]{figure/dir-com/head_orientation_pdf_pitch.csv}\headOrientationPdfPitch
\pgfplotstableread[col sep=comma]{figure/dir-com/head_angular_velocity_percentiles.csv}\headAngVelPerc
\pgfplotstableread[col sep=comma]{figure/dir-com/head_angular_velocity_vs_window_size.csv}\meanAngularVelocity

\pgfplotstableread[col sep=comma]{figure/dir-com/hand_planar_distance_pdf.csv}\handPlanarDistancePdf
\pgfplotstableread[col sep=comma]{figure/dir-com/hand_vertical_distance_percentiles.csv}\handVerticalDistancePercentiles
\pgfplotstableread[col sep=comma]{figure/dir-com/hand_persistence_pdf.csv}\handPersistencePdf

\title{MmWave for Extended Reality: Open User Mobility Dataset, Characterisation, and Impact on Link Quality}

\author{
Alexander Marin\v{s}ek~\orcidlink{0000-0001-9696-5365},
Sam De Kunst~\orcidlink{0009-0004-6096-0967},
Gilles Callebaut~\orcidlink{0000-0003-2413-986X},
Lieven De Strycker~\orcidlink{0000-0001-8172-9650},
Liesbet Van der Perre~\orcidlink{0000-0002-9158-9628}
\thanks{All authors are with ESAT-WaveCore, KU Leuven, 9000 Ghent, Belgium. Corresponding author: alexander.marinsek@kuleuven.be
}}

\narxiv{
\markboth{IEEE Communications Magazine,~Vol.~X, No.~Y, June~2024}%
{How to Use the IEEEtran \LaTeX \ Templates}
}

\maketitle

\begin{abstract}
User mobility in \gls{xr} can have a major impact on \gls{mmw} links and may require dedicated mitigation strategies to ensure reliable connections and avoid outage. The available prior art has predominantly focused on \gls{xr} applications with constrained user mobility and limited impact on \gls{mmw} channels. We have performed dedicated experiments to extend the characterisation of relevant future \gls{xr} use cases featuring a high degree of user mobility. To this end, we have carried out a tailor\nobreakdash-made measurement campaign and conducted a characterisation of the collected tracking data, including the approximation of the data using statistical distributions. Moreover, we have provided an interpretation of the possible impact of the recorded mobility on \gls{mmw} technology. The dataset is made publicly accessible to provide a testing ground for wireless system design and to enable further \gls{xr} mobility modelling.
\end{abstract}

\begin{IEEEkeywords}
Extended reality, head-mounted display, mobility, wireless, millimeter-wave
\end{IEEEkeywords}

\glsresetall

\section{Introduction}

\begin{figure}[h]
    \centering
    \plotFig{\plotFlag}{\input{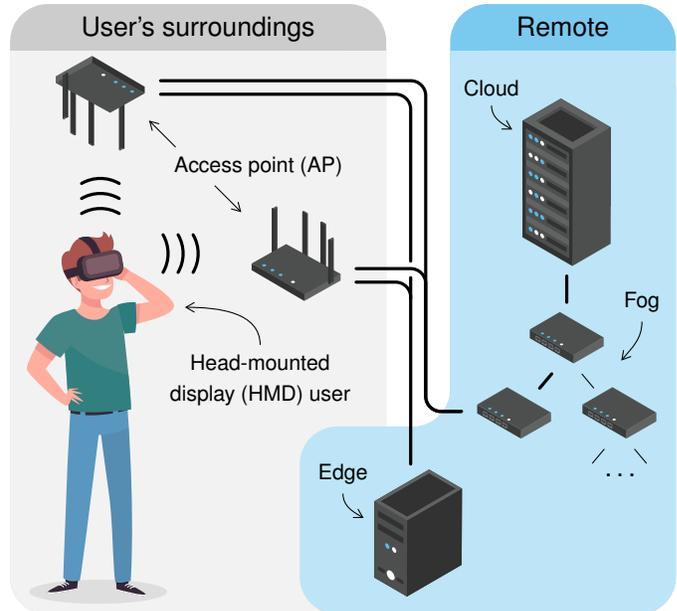}}
    \caption{Wireless remote rendering, which can take place in the network's edge, fog, or cloud. In a cellular network, the \glspl{ap} would be substituted by base stations. 
    }
    \label{fig:xr_architecture}
\end{figure}

\Gls{mmw} technology has been coined the great enabler of wireless \gls{xr} \narxiv{\cite{abari_enabling_2017, bastug_toward_2017, blandino_head_2021, chukhno_interplay_2022, morin_toward_2022}}\arxiv{\cite{abari_enabling_2017, bastug_toward_2017, blandino_head_2021, chukhno_interplay_2022, morin_toward_2022, giordani_toward_2020}}. This is primarily due to the multi\nobreakdash-\SI{}{\giga\bps} data rates that \gls{mmw} technology can provide, as a consequence of the large bandwidth availability in the \gls{mmw} spectrum (\SIrange{30}{300}{\giga\hertz}). Hence, \gls{mmw} links are distinguished by their real\nobreakdash-time high\nobreakdash-throughput streaming capabilities. Illustrated in \Cref{fig:xr_architecture}, \gls{mmw} links can connect a \gls{hmd} to remote processing infrastructure in the edge, fog, or cloud, where high\nobreakdash-fidelity \gls{xr} content rendering takes place~\cite{bastug_toward_2017, morin_toward_2022}. Thus, remote rendering and \gls{mmw} technology can together enable high video quality without compromising on freedom of movement, such as in the case of tethered connections. Moreover, remote rendering can reduce \gls{hmd} hardware complexity and, therefore, aid ergonomics and affordability. 

\Gls{mmw} systems are known for being highly directional by design, while the \gls{mmw} links are notorious for their relatively low robustness~\cite{cai_dynamic_2020, marinsek_impact_2023}. This can result in high losses when the radiation patterns of two communicating devices are misaligned or the link between them becomes obstructed~\cite{saha_fast_2018, gustafson_characterization_2012}. Equipping \glspl{hmd} with \gls{mmw} technology further magnifies the problem, owing to the high degree of movement exhibited by \gls{hmd} users~\cite{lincoln_low_2017}. \emph{
It is the understanding of this user mobility that holds a significant part of the key to deploying and configuring \gls{mmw} networks with support for \gls{xr} applications.}

User mobility, its impact on \gls{mmw} links, and potential mitigation strategies have been the subjects of past research~\cite{lincoln_low_2017, corbillon_360-degree_2017, blandino_head_2021, dong_why_2022, abari_enabling_2017, chukhno_interplay_2022, abari_enabling_2017}. However, only a subset of \gls{xr} use cases, exhibiting limited user mobility, have been considered in the prior art. In this work, we focus on relevant future \gls{xr} applications featuring a high degree of user mobility that could negatively impact wireless \gls{mmw} links. The target applications namely encourage users to explore their environment in all \gls{6dof} and to conduct rapid movements. That is, users both move in a 3\nobreakdash-dimensional space and rotate around three rotational axes, as shown in \Cref{fig:measurement_environment}. To facilitate this, we have carried out a tailor\nobreakdash-made measurement campaign, allowing us to characterise user movement in high\nobreakdash-mobility \gls{xr} use cases. Furthermore, we elaborate on the characterisation results to assess the possible impact of mobility on \gls{mmw} links. \emph{By doing so, we provide the following contributions:}
\begin{itemize}\vspace{.5em}\setlength\itemsep{.5em}
\item \emph{A publicly accessible dataset~\cite{cacerumd_zenodo}}, consisting of user tracking data for the head, hands, and body in \gls{6dof} with a total duration of \SI{45}{\hour}. This also includes \glspl{ssq} and reference measurements for evaluating tracking system accuracy.
\item \emph{A characterisation and statistical model\nobreakdash-based distribution} approximation of user mobility in a science\nobreakdash-fiction role\nobreakdash-playing game, an immersive training application, and an action\nobreakdash-packed shooter. 
\item \emph{An evaluation of the possible impact of user mobility on wireless \gls{mmw} links}, emphasizing the different adversities that comprise mobility.
\end{itemize}\vspace{.5em}

The article first describes the measurement campaign and places it alongside prior art in \Cref{sec:meas_campaign}. The collected tracking data are characterised in \Cref{sec:mobility_characterisation}, while \Cref{sec:impact_on_mmwave} assesses the possible impact of mobility on \gls{mmw} links. Finally, \Cref{sec:conclusion} summarises the key findings and outlines future research prospects.

\section{Measurement campaign and mobility dataset}
\label{sec:meas_campaign}

\narxiv{Several publicly\nobreakdash-accessible \gls{hmd} user mobility datasets and models are readily available at the time of writing~\cite{corbillon_360-degree_2017, blandino_head_2021,chukhno_interplay_2022, dong_why_2022}. 
These include \SI{360}{\degree} cinematic experiences~\cite{corbillon_360-degree_2017}, virtual museum visits~\cite{blandino_head_2021}, and pedestrian dynamics~\cite{chukhno_interplay_2022}. }%
\arxiv{Several publicly\nobreakdash-accessible \gls{hmd} user mobility datasets and models are readily available at the time of writing~\cite{corbillon_360-degree_2017, lo_360_2017, faye_open_2017, yan_ridi_2018, fremerey_avtrack360_2018, emery_openendedvr_2021, jin_where_2022, chakareski_6dof_2020, blandino_head_2021, struye_generating_2022, chukhno_interplay_2022, dong_why_2022, dong_why_2022, wei_6dof_2023}, summarised in \Cref{tab:dataset_and_model}. These include \SI{360}{\degree} cinematic experiences~\cite{corbillon_360-degree_2017, lo_360_2017, fremerey_avtrack360_2018, jin_where_2022}, virtual museum visits~\cite{chakareski_6dof_2020, blandino_head_2021, struye_generating_2022}, daily activities~\cite{faye_open_2017, yan_ridi_2018, emery_openendedvr_2021, wei_6dof_2023}, gaming~\cite{dong_why_2022}, and pedestrian dynamics~\cite{chukhno_interplay_2022}. }%
However, the aforementioned feature limited user mobility, which is insufficient for assessing the possible more severe impact of mobility on \gls{mmw} links. On the other hand, the prior art that includes relevant high\nobreakdash-mobility \gls{xr} use cases captures exclusively the orientation of the user's head in 3DoF~\cite{dong_why_2022}. The resulting tracking data only permits the analysis of angular head dynamics. To this end, we have conducted a measurement campaign and captured \gls{hmd}, handheld controller, and body tracker position and orientation in \gls{xr} applications requiring a high degree of user movement. To the best of our knowledge, this is the first publicly accessible dataset to include \gls{6dof} tracking for multiple body parts in high\nobreakdash-mobility scenarios.

\arxiv{
\begin{table*}[t]
\centering
\caption{An overview of existing mobility datasets and models. The \textbf{DoF} column lists the number of degrees of freedom in the collected or generated data, column \textbf{Type} indicates whether the tracking data originate from measurements or a mobility model, the \textbf{System} column briefly lists the employed measurement hardware or algorithm in the case of mobility models, the \textbf{Data} column indicates the provided data types (\gls{hmd} refers to tracking data for the \gls{hmd}), the \textbf{Use case} column indicates the activity during which data were recorded or for which they were generated, the \textbf{Duration} column lists the approximate total duration of the recordings or generated tracking data (model), while the sampling frequency is listed in the last column.}
\label{tab:dataset_and_model}
\resizebox{\linewidth}{!}{


\begin{tabular}{|r|ccllllrr|}
    
    \toprule
    
    \textbf{(et al.)} & \textbf{Year} & \textbf{DoF} & \textbf{Type} & \textbf{System} & \textbf{Data} & \textbf{Use case} & \textbf{Duration (h)} & \textbf{Freq (Hz)} \\
    
    \midrule
        
    Corbillon    \cite{corbillon_360-degree_2017}& \small2017 & 3 & Data & Razer OSVR & HMD & \qty{360}{\degree} video & \SI{6}{} & \SI{30}{}  \\
    
    \rowcolor{lightgray} Lo           \cite{lo_360_2017}              & \small2017 & 3 & Data & Oculus Rift DK2 & HMD, video & \qty{360}{\degree} video & \SI{8}{} & \SI{30}{} \\
    
    Faye         \cite{faye_open_2017}           & \small2017 & 3 & Data & \makecell[l]{Nexus 5X phone \\ LG Watch \\ Smartglasses} & \makecell[l]{GPS, steps \\ Heart rate, steps \\ Gaze, Velocity} & Daily activities & \SI{60}{} & \SI{100}{} \\
    
    \rowcolor{lightgray} Yan         \cite{yan_ridi_2018}        & \small2018 & 6 & Data & Phone IMU & IMU sensors & Phone on body & \SI{100}{} & \SI{200}{}  \\
    
    Fremery      \cite{fremerey_avtrack360_2018} & \small2018 & 3 & Data & HTC Vive & HMD & \qty{360}{\degree} video & \SI{8}{} & \SI{200}{} \\
    
    \rowcolor{lightgray} Chakareski   \cite{chakareski_6dof_2020}     & \small2020 & 6 & Data & HTC Vive Wireless & HMD & Virtual museum & $<$\SI{1}{} & \SI{250}{}  \\
    
    Emery     \cite{emery_openendedvr_2021}       & \small2021 & 3/6 & \makecell[l]{Data \\ Model} & \makecell[l]{Custom Oculus Rift \\ HMD + Eye gaze \\ Controllers} & \makecell[l]{HMD (3DoF) \\ Hands (6DoF) \\ Gaze, scenes} & Daily activities & \SI{4}{} & \SI{90}{}    \\
    
    \rowcolor{lightgray} Blandino     \cite{blandino_head_2021}       & \small2021 & 3 & Model & Extrapolation & HMD & Virtual museum & Variable & \SI{250}{}    \\
    
    Struye       \cite{struye_generating_2022}   & \small2022 & 3 & Model & Deep Learning & HMD & Virtual museum & Variable & \SI{17}{}    \\
    
    \rowcolor{lightgray} Dong     \cite{dong_why_2022}       & \small2022 & 3 & Data & Oculus Quest 2 & HMD, Heart, SSQ &  Gaming & \SI{13}{} & \SI{250}{}    \\
                
    Jin          \cite{jin_where_2022}           & \small2022 & 3 & Data & HTC Vive Pro Eye & HMD, Gaze & \qty{360}{\degree} video & \SI{54}{} & \SI{120}{} \\
                
    \rowcolor{lightgray} Chukhno \cite{chukhno_interplay_2022} & \small2022 & 3 & Model & Particle statistics & HMD & Pedestrian flow \cite{farina_walking_2017} & Variable & Variable \\
    
    Wei          \cite{wei_6dof_2023}           & \small2023 & 6 & Data & \makecell[l]{HTC Vive Pro Eye \\ Vive Controllers \\ RGB Camera} & \makecell[l]{HMD, Gaze \\ Hand controls \\ Video} & Daily activities & \SI{3}{} & \SI{50}{} \\
    
    \midrule
    
    \textit{Ours} & \textit{\small2023} & \textit{6} & \textit{Data} & \makecell[l]{\textit{HTC Vive Pro 2} \\ \textit{Vive Controllers} \\ \textit{Tundra tracker} } & \makecell[l]{\textit{HMD, SSQ} \\ \textit{Hand controls} \\ \textit{Waist tracking}} & \makecell[l]{\textit{Gaming} \\ \textit{Training}} & \textit{\SI{45}{}} & \makecell[l]{~\\ \textit{\SI{250}{}} \\ ~ } \\
    
    \bottomrule
    
\end{tabular}


}
\end{table*}
}

\subsection{Measurement setup}
\label{sec:meas_campaign:measurement_setup}

\begin{figure}[h]
    \centering
    \plotFig{\plotFlag}{\input{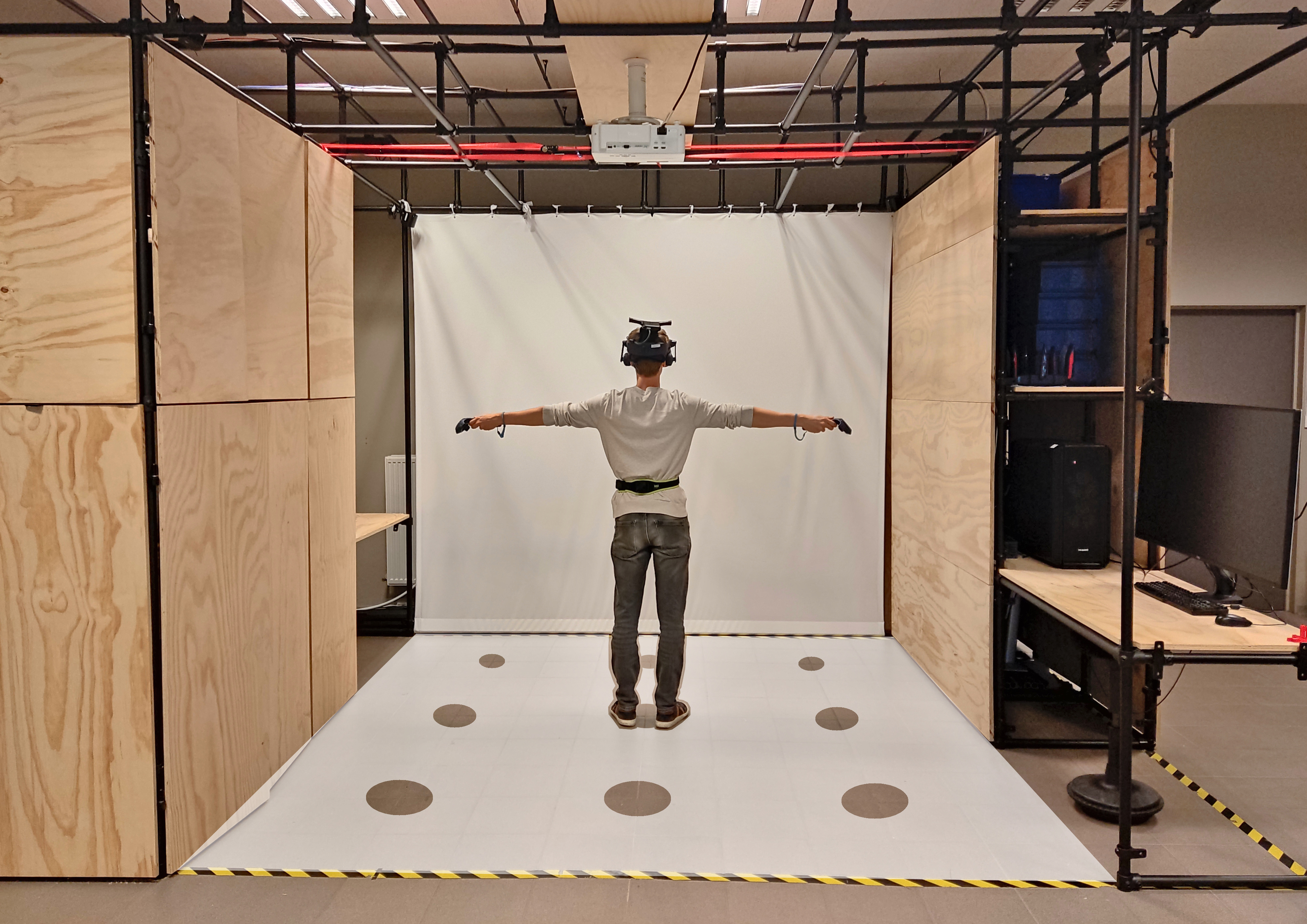}}
    \vspace{-3mm}
    \caption{Measurement environment and a volunteer conducting the T-Pose. The \gls{ir} beacons are highlighted in blue, while the tracked \gls{xr} devices are marked in red. The axes left of the \gls{hmd} show the \glsposs{hmd} body frame (orientation), with positive rotations determined by the right-hand rule. The axes at the bottom right of the picture show the Cartesian coordinate system (position). The $3\times3\,$\SI{}{\meter} playspace is highlighted white and circles mark the reference measurement locations.}
    \label{fig:measurement_environment}
\end{figure}

\Cref{fig:measurement_environment} shows the measurement setup and highlights its key components.
\narxiv{We employed the HTC Vive Pro 2 \gls{hmd}, along with two handheld controllers, and a body tracker, to sample mobility data for the head, hands, and body.}\arxiv{We collected \gls{6dof} tracking data for the below four body parts:
\begin{itemize}\vspace{.25em}\setlength\itemsep{.25em}
\item The head, using the HTC Vive Pro~2 \gls{hmd}~\cite{htc_vive_pro2}.
\item Both hands, with handheld controllers~\cite{vive_controller}. 
\item The waist, using a Tundra Labs body tracker~\cite{tundra_tracker}. 
\end{itemize}\vspace{.25em}}
The \gls{xr} devices position themselves using on\nobreakdash-board \glspl{imu} and external \gls{ir} beacons. 
The latter sweep the horizontal and vertical angular domain using \gls{ir} beams\arxiv{\xspace with a frequency of \SI{60}{\hertz}} to provide precise tracking\arxiv{~\cite{valve_lighthouse}}. We used four \gls{ir} beacons to ensure a high degree of positioning accuracy and to avoid tracking outage, where the latter could negatively impact tracking accuracy~\cite{niehorster_accuracy_2017}. Tracking data is sampled at a \SI{500}{\hertz} frequency on the \gls{hmd} and at \SI{250}{\hertz} on the handheld controllers and body tracker.
\arxiv{To extract user tracking information, we employed the Brekel OpenVR Recorder~\cite{brekel_openvr} and ran it as a background process. }%

\subsection{Experiment procedure}

\arxiv{
\Cref{fig:experiment_procedure} shows the procedure applied for each volunteer, with the illustrated steps outlined in the bellow paragraphs.
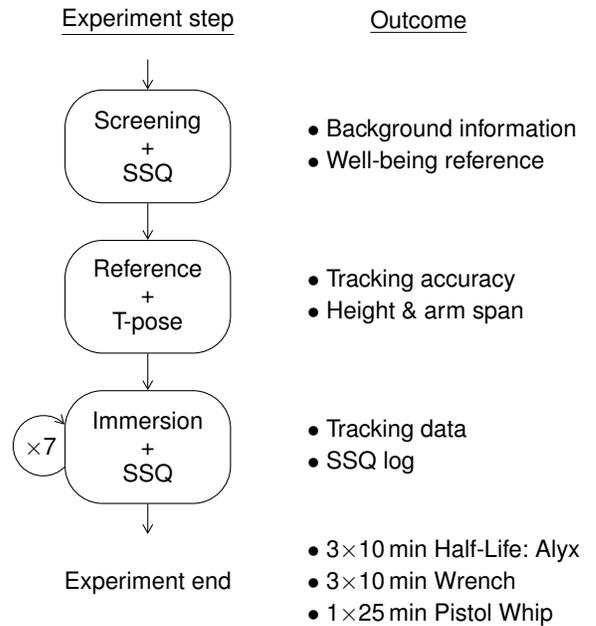
\begin{figure}[h]
    \centering
    \plotFig{\plotFlag}{\begin{tikzpicture}[
    font=\sffamily\small,
    roundrec/.style={
        anchor=center, 
        align=center,
        draw=black, 
        rounded corners=0.5cm,
        minimum width=2.2cm, 
        minimum height=1.5cm, 
    }
]

\node[] (zero) at (0, 1.65) {\underline{Experiment step}};
\node[anchor=center, align=center] at (3.6, 1.65) {\underline{Outcome}};

\node[roundrec] (one) at (0, 0) {Screening\\+\\SSQ};
\node[anchor=west, align=left] at (2, 0) {
    $\bullet$ Background information \\[2pt]
    $\bullet$ Well-being reference
};

\node[roundrec] (two) at (0,-2) {Reference\\+\\T-pose};
\node[anchor=west, align=left] at (2, -2) {
    $\bullet$ Tracking accuracy \\[2pt]
    $\bullet$ Height \& arm span
};

\node[roundrec] (three) at (0,-4) {Immersion\\+\\SSQ};
\node[anchor=west, align=left] at (2, -4) {
    $\bullet$ Tracking data \\[2pt]
    $\bullet$ SSQ log
};
\draw[black, {angle 60}-] ([yshift=3mm]three.west) arc (45:315:4mm);
\node[] at ([xshift=-3.2mm]three.west) {$\times$7};

\node[] (four) at (0, -5.8) {Experiment end};
\node[anchor=west, align=left] at (2, -5.8) {
    $\bullet$ 3$\times$10$\,$min \hali \\[2pt]
    $\bullet$ 3$\times$10$\,$min \wren \\[2pt]
    $\bullet$ 1$\times$25$\,$min \piwi
};

\draw[black, -{angle 60}] ([yshift=4mm]one.north) -- (one.north);
\draw[black, -{angle 60}] (one.south) -- (two.north);
\draw[black, -{angle 60}] (two.south) -- (three.north);
\draw[black, -{angle 60}] (three.south) -- ([yshift=-4mm]three.south);
        
\end{tikzpicture}  }
    \caption{Experiment procedure per volunteer. The last step (block) is repeated seven times in order to collect the seven tracking data recordings, listed at the bottom right.}
    \label{fig:experiment_procedure}
\end{figure}
}

\narxiv{
A total of 33~volunteers, aged \SIrange{19}{40}{years}, have together contributed to \SI{45}{\hour} of tracking data recordings. Volunteers were given a screening questionnaire before the experiment, where they provided informed consent for the further processing of their data, while disclosing their basic information and experience with \gls{xr}. }%
\arxiv{\subsubsection{Screening and initial SSQ}
A total of 33~volunteers have together contributed to \SI{45}{\hour} of tracking data recordings. Volunteers were given a screening questionnaire before the experiment, where they provided informed consent for the further processing of their data, while disclosing their age, gender, height, underlying medical conditions, participation in sports activities, and experience with \gls{xr}. The 11~female and 22~male volunteers were aged \SIrange{19}{40}{}~years, with the median and mean age, respectively, \SI{25}{} and \SI{26}{}~years. Following the screening, }%
\narxiv{Next, }the volunteers completed an initial \gls{ssq}, which serves as a reference of their well\nobreakdash-being before the experiment.

\arxiv{\subsubsection{Reference measurements and the T-pose}
Reference }%
\narxiv{Reference }measurements were made between experiments by placing the four devices on the floor at the positions illustrated in \Cref{fig:measurement_environment}. While the matching against floor markings is not exact due to limited precision when laying the devices on the floor, no position outliers were observed. We calculated that the spread of values around the mean position, the \gls{rmse}, for each of the 9~reference measurement positions, evaluated 33~times (once per volunteer), is less than \SI{5}{\milli\meter} for all four tracked \gls{xr} devices in the setup employed in the measurement campaign.
This was followed by the \emph{T\nobreakdash-pose} recording. Here, the participants stood at the centre of the playspace with their arms extended sideways, as shown in \Cref{fig:measurement_environment}.\arxiv{ We then recorded the tracking data for \SI{10}{\second}. This concluded the screening and reference measurement procedures.}

\arxiv{\subsubsection{Immersion and intermediate SSQs}
Volunteers }%
\narxiv{Volunteers }were immersed in three different applications, spending approximately \SIrange{20}{30}{\minute} in each. Breaks were scheduled every \SI{10}{\minute} in order for the volunteers to complete an \gls{ssq} and to freshen up. 
\narxiv{The three applications are:
\begin{itemize}\vspace{.25em}\setlength\itemsep{.25em}
\item \textbf{Half\nobreakdash-Life: Alyx}, a role\nobreakdash-playing game, set in a post\nobreakdash-apocalyptic world. The user traverses three different landscapes while solving puzzles and battling nemeses.
\item \textbf{Wrench}, an immersive car mechanic training application. The user is tasked with conducting three different maintenance operations on a car in a mechanic's workshop.
\item \textbf{Pistol Whip}, an action\nobreakdash-packed shooter. The user automatically moves along a linear trajectory in the game while dodging obstacles and shooting or punching villains.
\end{itemize}\vspace{.25em}}
\arxiv{The three applications are listed in \Cref{tab:applications}, with the observed user mobility characteristics listed alongside them. Our aim was to capture mobility profiles that are possibly adverse for wireless \gls{mmw} links. This includes \emph{ample movement}, for example, \SI{360}{\degree} head rotation and movement along the entire playspace. In addition, we were interested in \emph{fast movement}, such as rapid head rotation and swift lateral movement. After reviewing the applications in SteamVR and discussing with experienced \gls{xr} users, we foresaw that \wren would feature ample movement, while \piwi would result in fast movement. \hali was included since we estimated the in\nobreakdash-game puzzles and combat sequences would result in a mixture of both adverse movement types.}

\arxiv{
\begin{table*}[htbp]
    \centering
    \caption{Selected \gls{xr} applications for the experiment and the corresponding user mobility.}
    \label{tab:applications}
    
\begin{tabular}{lp{7.5cm}p{7.5cm}}
    \toprule
    \textbf{Application} & \textbf{Description} & \textbf{Mobility characteristics} \\
    \midrule
    \rowcolor{hl_gray} \textbf{Half-Life: Alyx} & A role-playing game set in a post-apocalyptic world. The user traverses three different landscapes while solving puzzles and battling nemeses. & Moderate movement as users explore their surroundings, with fast rotations during occasional high\nobreakdash-mobility action sequences. \\
    \midrule
    \rowcolor{wr_gray} \textbf{Wrench} & An immersive car mechanic training application. The user is tasked with conducting three different maintenance operations on a car in a mechanic's workshop. & Ample movement along the playspace and ample head rotation when working on a vehicle in a mechanic's garage. \\
    \midrule
    \rowcolor{pw_gray} \textbf{Pistol Whip} & An action-packed shooter. The user automatically moves along a linear trajectory in the game while dodging obstacles and shooting or punching villains. & Fast posture changes throughout the heated gunfight, with limited movement due to the automatic in\nobreakdash-game forwards trajectory. \\
    \bottomrule
\end{tabular}


\end{table*}
}

\begin{table*}[t]
    \centering
    \renewcommand{\arraystretch}{1.25}
    \caption{Mobility characterisation summary, showing the measured data percentiles P$_x$, where $x \in \{50, 98, 99, 100\}$, and the corresponding fitted distributions. Shades of gray represent the three applications, from darkest to lightest, respectively, Half-Life: Alyx, Wrench, and Pistol Whip. The letters in the ID column show the subsection within \Cref{sec:mobility_characterisation} where each variable is discussed. The percentiles highlighted with a bold dark blue font bear important information for \gls{mmw} link quality, discussed in \Cref{sec:impact_on_mmwave}. The vertical distance factor shows the complementary percentiles to highlight the lowest relative head position of users. The three distribution parameters, respectively, represent the shape or skewness, location along the abscissa, and the scaling factor. The min/max columns show the boundaries (inclusive, except for 0$<$), between which the distributions are defined. The rightmost column shows the results of the Kolmogorov-Smirnov goodness of fit evaluation.
    }
    \label{tab:mobility_characterisation}
    \resizebox{\linewidth}{!}{

\begin{tabular}{c|cl|rrrr|lrrr|rr|r}
    \toprule

    & \multicolumn{2}{c|}{} & \multicolumn{4}{c|}{\textbf{Measured percentiles}} & \multicolumn{7}{c}{\textbf{Fitted distributions}}\\
    
    & \textbf{ID} & \textbf{Quantity} & $\bm{\text{P}_{50}}$ & $\bm{\text{P}_{98}}$ & $\bm{\text{P}_{99}}$ & $\bm{\text{P}_{100}}$ & ~Name & Shape & Loc. & Scale & Min & Max & KS \\
    \midrule
    
    \parbox[t]{1mm}{\multirow{15}{*}{\rotatebox[origin=c]{90}{Head}}} 
    &\multirow{3}{*}{A}&\multirow{3}{*}{\makecell[l]{Planar distance (\SI{}{\meter})}} 
    &\cellcolor{hl_gray}\textbf{\color{RoyalPurple}0.3}&\cellcolor{hl_gray}1.1&\cellcolor{hl_gray}1.2&\cellcolor{hl_gray}2.0&
    \cellcolor{hl_gray}~Gamma&\cellcolor{hl_gray}2.06&\cellcolor{hl_gray}0&\cellcolor{hl_gray}0.19&\cellcolor{hl_gray}0$<$&\cellcolor{hl_gray}2.0&\cellcolor{hl_gray}0.02\\
    
    &&&\cellcolor{wr_gray}\textbf{\color{RoyalPurple}0.4}&\cellcolor{wr_gray}1.2&\cellcolor{wr_gray}1.4&\cellcolor{wr_gray}2.3&
    \cellcolor{wr_gray}~Gamma&\cellcolor{wr_gray}2.54&\cellcolor{wr_gray}0&\cellcolor{wr_gray}0.19&\cellcolor{wr_gray}0$<$&\cellcolor{wr_gray}2.3&\cellcolor{wr_gray}0.02\\
    
    &&&\cellcolor{pw_gray}\textbf{\color{RoyalPurple}0.3}&\cellcolor{pw_gray}0.9&\cellcolor{pw_gray}1.0&\cellcolor{pw_gray}1.4&
    \cellcolor{pw_gray}~Gamma&\cellcolor{pw_gray}2.33&\cellcolor{pw_gray}0&\cellcolor{pw_gray}0.15&\cellcolor{pw_gray}0$<$&\cellcolor{pw_gray}1.4&\cellcolor{pw_gray}0.02\\

    \cline{2-14}
    
    &\multirow{3}{*}{B}&\multirow{3}{*}{\makecell[l]{Vertical distance factor\\[0.25em]{\fontsize{7}{6}\selectfont \textit{Complementary percentiles}}}} 
    &\cellcolor{hl_gray}1.0&\cellcolor{hl_gray}0.5&\cellcolor{hl_gray}0.4&\cellcolor{hl_gray}\textbf{\color{RoyalPurple}0.3}&
    \cellcolor{hl_gray}~Skewed~Cauchy&\cellcolor{hl_gray}-0.54&\cellcolor{hl_gray}0.99&\cellcolor{hl_gray}0.02&\cellcolor{hl_gray}0.3&\cellcolor{hl_gray}1.17&\cellcolor{hl_gray}0.04\\
    
    &&&\cellcolor{wr_gray}0.9&\cellcolor{wr_gray}0.5&\cellcolor{wr_gray}0.3&\cellcolor{wr_gray}\textbf{\color{RoyalPurple}0.2}&
    \cellcolor{wr_gray}~Skewed~Cauchy&\cellcolor{wr_gray}-0.57&\cellcolor{wr_gray}0.98&\cellcolor{wr_gray}0.04&\cellcolor{wr_gray}0.2&\cellcolor{wr_gray}1.12&\cellcolor{wr_gray}0.04\\
    
    &&&\cellcolor{pw_gray}1.0&\cellcolor{pw_gray}0.5&\cellcolor{pw_gray}0.5&\cellcolor{pw_gray}\textbf{\color{RoyalPurple}0.3}&
    \cellcolor{pw_gray}~Skewed~Cauchy&\cellcolor{pw_gray}-0.76&\cellcolor{pw_gray}1.00&\cellcolor{pw_gray}0.02&\cellcolor{pw_gray}0.3&\cellcolor{pw_gray}1.18&\cellcolor{pw_gray}0.03\\

    \cline{2-14}
    
    &\multirow{3}{*}{C}&\multirow{3}{*}{\makecell[l]{Pitch orientation (\SI{}{\degree})}} 
    &\cellcolor{hl_gray}\textbf{\color{RoyalPurple}20}&\cellcolor{hl_gray}70&\cellcolor{hl_gray}75&\cellcolor{hl_gray}90&
    \cellcolor{hl_gray}~Skewed~normal&\cellcolor{hl_gray}1.98&\cellcolor{hl_gray}2.96&\cellcolor{hl_gray}28.23&\cellcolor{hl_gray}-80&\cellcolor{hl_gray}90&\cellcolor{hl_gray}0.03\\
    
    &&&\cellcolor{wr_gray}\textbf{\color{RoyalPurple}32}&\cellcolor{wr_gray}74&\cellcolor{wr_gray}79&\cellcolor{wr_gray}90&
    \cellcolor{wr_gray}~Skewed~normal&\cellcolor{wr_gray}-2.96&\cellcolor{wr_gray}59.32&\cellcolor{wr_gray}41.04&\cellcolor{wr_gray}-82&\cellcolor{wr_gray}90&\cellcolor{wr_gray}0.02\\
    
    &&&\cellcolor{pw_gray}3&\cellcolor{pw_gray}33&\cellcolor{pw_gray}44&\cellcolor{pw_gray}90&
    \cellcolor{pw_gray}~Logistic&\cellcolor{pw_gray}/&\cellcolor{pw_gray}2.85&\cellcolor{pw_gray}5.93&\cellcolor{pw_gray}-70&\cellcolor{pw_gray}90&\cellcolor{pw_gray}0.03\\

    \cline{2-14}
    
    &\multirow{3}{*}{D}&\multirow{3}{*}{\makecell[l]{Lateral velocity (\SI{}{\meter\per\second})}} 
    &\cellcolor{hl_gray}0.1&\cellcolor{hl_gray}0.6&\cellcolor{hl_gray}0.8&\cellcolor{hl_gray}3.2&
    \cellcolor{hl_gray}~Weibull&\cellcolor{hl_gray}0.93&\cellcolor{hl_gray}0&\cellcolor{hl_gray}0.13&\cellcolor{hl_gray}0$<$&\cellcolor{hl_gray}3.2&\cellcolor{hl_gray}0.04\\
    
    &&&\cellcolor{wr_gray}0.1&\cellcolor{wr_gray}0.7&\cellcolor{wr_gray}0.9&\cellcolor{wr_gray}2.4&
    \cellcolor{wr_gray}~Weibull&\cellcolor{wr_gray}0.80&\cellcolor{wr_gray}0&\cellcolor{wr_gray}0.12&\cellcolor{wr_gray}0$<$&\cellcolor{wr_gray}2.4&\cellcolor{wr_gray}0.03\\
    
    &&&\cellcolor{pw_gray}0.2&\cellcolor{pw_gray}1.4&\cellcolor{pw_gray}1.6&\cellcolor{pw_gray}3.2&
    \cellcolor{pw_gray}~Weibull&\cellcolor{pw_gray}0.85&\cellcolor{pw_gray}0&\cellcolor{pw_gray}0.29&\cellcolor{pw_gray}0$<$&\cellcolor{pw_gray}3.2&\cellcolor{pw_gray}0.05\\

    \cline{2-14}
    
    &\multirow{3}{*}{E}&\multirow{3}{*}{\makecell[l]{Angular velocity (\SI{}{\degree\per\second})}} 
    &\cellcolor{hl_gray}15&\cellcolor{hl_gray}180&\cellcolor{hl_gray}224&\cellcolor{hl_gray}\textbf{\color{RoyalPurple}646}&
    \cellcolor{hl_gray}~Log-normal&\cellcolor{hl_gray}1.27&\cellcolor{hl_gray}0&\cellcolor{hl_gray}15.84&\cellcolor{hl_gray}0$<$&\cellcolor{hl_gray}646&\cellcolor{hl_gray}0.01\\
    
    &&&\cellcolor{wr_gray}10&\cellcolor{wr_gray}158&\cellcolor{wr_gray}197&\cellcolor{wr_gray}\textbf{\color{RoyalPurple}550}&
    \cellcolor{wr_gray}~Log-normal&\cellcolor{wr_gray}1.38&\cellcolor{wr_gray}0&\cellcolor{wr_gray}10.20&\cellcolor{wr_gray}0$<$&\cellcolor{wr_gray}550&\cellcolor{wr_gray}0.01\\
    
    &&&\cellcolor{pw_gray}21&\cellcolor{pw_gray}166&\cellcolor{pw_gray}208&\cellcolor{pw_gray}\textbf{\color{RoyalPurple}549}&
    \cellcolor{pw_gray}~Log-normal&\cellcolor{pw_gray}1.04&\cellcolor{pw_gray}0&\cellcolor{pw_gray}20.91&\cellcolor{pw_gray}0$<$&\cellcolor{pw_gray}549&\cellcolor{pw_gray}0.01\\

    \midrule

    \parbox[t]{1mm}{\multirow{12}{*}{\rotatebox[origin=c]{90}{Hands}}} 
    &\multirow{3}{*}{F}&\multirow{3}{*}{\makecell[l]{Planar distance (\SI{}{\meter})}} 
    &\cellcolor{hl_gray}\textbf{\color{RoyalPurple}0.2}&\cellcolor{hl_gray}0.6&\cellcolor{hl_gray}0.6&\cellcolor{hl_gray}1.0&
    \cellcolor{hl_gray}~Log-normal&\cellcolor{hl_gray}0.32&\cellcolor{hl_gray}0&\cellcolor{hl_gray}0.33&\cellcolor{hl_gray}0$<$&\cellcolor{hl_gray}1&\cellcolor{hl_gray}0.04\\
    
    &&&\cellcolor{wr_gray}\textbf{\color{RoyalPurple}0.3}&\cellcolor{wr_gray}0.5&\cellcolor{wr_gray}0.6&\cellcolor{wr_gray}1.0&
    \cellcolor{wr_gray}~Log-normal&\cellcolor{wr_gray}0.22&\cellcolor{wr_gray}0&\cellcolor{wr_gray}0.48&\cellcolor{wr_gray}0$<$&\cellcolor{wr_gray}1&\cellcolor{wr_gray}0.01\\
    
    &&&\cellcolor{pw_gray}\textbf{\color{RoyalPurple}0.3}&\cellcolor{pw_gray}0.6&\cellcolor{pw_gray}0.7&\cellcolor{pw_gray}1.0&
    \cellcolor{pw_gray}~Log-normal&\cellcolor{pw_gray}0.29&\cellcolor{pw_gray}0&\cellcolor{pw_gray}0.44&\cellcolor{pw_gray}0$<$&\cellcolor{pw_gray}1&\cellcolor{pw_gray}0.03\\

    \cline{2-14}
    
    &\multirow{3}{*}{G}&\multirow{3}{*}{\makecell[l]{Vertical distance (\SI{}{\meter})}} 
    &\cellcolor{hl_gray}-0.4&\cellcolor{hl_gray}\textbf{\color{RoyalPurple}0.0}&\cellcolor{hl_gray}0.1&\cellcolor{hl_gray}0.6&
    \cellcolor{hl_gray}~Weibull&\cellcolor{hl_gray}3.91&\cellcolor{hl_gray}-1.17&\cellcolor{hl_gray}0.82&\cellcolor{hl_gray}-1.1&\cellcolor{hl_gray}0.6&\cellcolor{hl_gray}0.05\\
    
    &&&\cellcolor{wr_gray}-0.4&\cellcolor{wr_gray}\textbf{\color{RoyalPurple}0.1}&\cellcolor{wr_gray}0.2&\cellcolor{wr_gray}0.6&
    \cellcolor{wr_gray}~Logistic&\cellcolor{wr_gray}/&\cellcolor{wr_gray}-0.42&\cellcolor{wr_gray}0.13&\cellcolor{wr_gray}-1.1&\cellcolor{wr_gray}0.6&\cellcolor{wr_gray}0.01\\ 
    
    &&&\cellcolor{pw_gray}-0.4&\cellcolor{pw_gray}\textbf{\color{RoyalPurple}0.1}&\cellcolor{pw_gray}0.1&\cellcolor{pw_gray}0.6&
    \cellcolor{pw_gray}~Weibull&\cellcolor{pw_gray}4.45&\cellcolor{pw_gray}-1.51&\cellcolor{pw_gray}1.20&\cellcolor{pw_gray}-1.1&\cellcolor{pw_gray}0.6&\cellcolor{pw_gray}0.06\\

    \cline{2-14}
    
    &\multirow{3}{*}{H}&\multirow{3}{*}{\makecell[l]{Above head duration (\SI{}{\second})}} 
    &\cellcolor{hl_gray}0.5&\cellcolor{hl_gray}5.4&\cellcolor{hl_gray}7.3&\cellcolor{hl_gray}\textbf{\color{RoyalPurple}22.7}&
    \cellcolor{hl_gray}~Weibull&\cellcolor{hl_gray}0.91&\cellcolor{hl_gray}0&\cellcolor{hl_gray}0.94&\cellcolor{hl_gray}0$<$&\cellcolor{hl_gray}22.7&\cellcolor{hl_gray}0.08\\
    
    &&&\cellcolor{wr_gray}1.3&\cellcolor{wr_gray}12.6&\cellcolor{wr_gray}16.6&\cellcolor{wr_gray}\textbf{\color{RoyalPurple}46.7}&
    \cellcolor{wr_gray}~Weibull&\cellcolor{wr_gray}0.82&\cellcolor{wr_gray}0&\cellcolor{wr_gray}2.14&\cellcolor{wr_gray}0$<$&\cellcolor{wr_gray}46.7&\cellcolor{wr_gray}0.03\\
    
    &&&\cellcolor{pw_gray}0.4&\cellcolor{pw_gray}2.1&\cellcolor{pw_gray}2.6&\cellcolor{pw_gray}\textbf{\color{RoyalPurple}22.3}&
    \cellcolor{pw_gray}~Weibull&\cellcolor{pw_gray}1.16&\cellcolor{pw_gray}0&\cellcolor{pw_gray}0.62&\cellcolor{pw_gray}0$<$&\cellcolor{pw_gray}22.3&\cellcolor{pw_gray}0.04\\

    \cline{2-14}
    
    &\multirow{3}{*}{I}&\multirow{3}{*}{\makecell[l]{Lateral velocity (\SI{}{\meter\per\second})}} 
    &\cellcolor{hl_gray}0.1&\cellcolor{hl_gray}1.3&\cellcolor{hl_gray}1.6&\cellcolor{hl_gray}9.7&
    \cellcolor{hl_gray}~Weibull&\cellcolor{hl_gray}0.84&\cellcolor{hl_gray}0&\cellcolor{hl_gray}0.23&\cellcolor{hl_gray}0$<$&\cellcolor{hl_gray}9.7&\cellcolor{hl_gray}0.04\\
    
    &&&\cellcolor{wr_gray}0.1&\cellcolor{wr_gray}1.2&\cellcolor{wr_gray}1.4&\cellcolor{wr_gray}9.1&
    \cellcolor{wr_gray}~Weibull&\cellcolor{wr_gray}0.80&\cellcolor{wr_gray}0&\cellcolor{wr_gray}0.20&\cellcolor{wr_gray}0$<$&\cellcolor{wr_gray}9.1&\cellcolor{wr_gray}0.05\\
    
    &&&\cellcolor{pw_gray}0.4&\cellcolor{pw_gray}3.2&\cellcolor{pw_gray}3.9&\cellcolor{pw_gray}9.9&
    \cellcolor{pw_gray}~Weibull&\cellcolor{pw_gray}0.86&\cellcolor{pw_gray}0&\cellcolor{pw_gray}0.63&\cellcolor{pw_gray}0$<$&\cellcolor{pw_gray}9.9&\cellcolor{pw_gray}0.02\\

    \bottomrule
\end{tabular}
}

\end{table*}

\subsection{High-Mobility 6DoF Tracking Dataset}

\arxiv{
\Cref{fig:post_processing_procedure} shows the applied post\nobreakdash-processing procedures, outlined in the bellow paragraphs, and lists the four building blocks of the generated dataset, accessible at~\cite{cacerumd_zenodo}.
\begin{figure}[h]
    \centering
    \plotFig{\plotFlag}{\begin{tikzpicture}[
    font=\sffamily\small,
    roundrec/.style={
        anchor=center, 
        align=center,
        draw=black, 
        rounded corners=0.5cm,
        minimum width=2.75cm, 
        minimum height=1.4cm, 
    }
]

\node[] (zero) at (0, 1.6) {\underline{Processing step}};
\node[anchor=west, align=center] at (2, 1.6) {\underline{Reason / threshold / technique}};

\node[roundrec] (one) at (0, 0) {Segment removal\\[2pt](manual)};
\node[anchor=west, align=left] at (2, 0) {
    $\bullet$ Water break \\[2pt]
    $\bullet$ Equipment re-adjusting
};

\node[roundrec] (two) at (0,-2) {Outlier removal\\[2pt](automated)};
\node[anchor=west, align=left] at (2, -2) {
    $\bullet$ \gls{hmd} $<\,$800$\,^{\circ}$/s$^2$ \\[2pt]
    $\bullet$ Handheld controller $<\,$10$\,$m/s \\[2pt]
    $\bullet$ Body tracker $<\,$10$\,$m/s
};

\node[roundrec] (three) at (0,-4) {Resampling};
\node[anchor=west, align=left] at (2, -4) {
    $\bullet$ 250$\,$Hz \\[2pt]
    $\bullet$ (Spherical) linear interpolation
};

\draw[black, -{angle 60}] ([yshift=4mm]one.north) -- (one.north);
\draw[black, -{angle 60}] (one.south) -- (two.north);
\draw[black, -{angle 60}] (two.south) -- (three.north);
\draw[black, -{angle 60}] (three.south) -- ([yshift=-4mm]three.south);

\draw[black, dashed] (-2,-5.5) -- (6.8,-5.5) node[midway, text opacity=1, fill=white, opacity=1] {Resulting dataset};

\node[roundrec, anchor=west, minimum width=2cm] at (-2,-6.6) {Reference\\[2pt]\& T-pose};
\node[roundrec, anchor=west, minimum width=2cm] at (0.2,-6.6) {Tracking};
\node[roundrec, anchor=west, minimum width=2cm] at (2.4,-6.6) {Controller\\[2pt]input};
\node[roundrec, anchor=west, minimum width=2cm] at (4.6,-6.6) {SSQ};

\end{tikzpicture}  }
    \caption{Post-processing procedure with the resulting dataset elements at the bottom.}
    \label{fig:post_processing_procedure}
\end{figure}
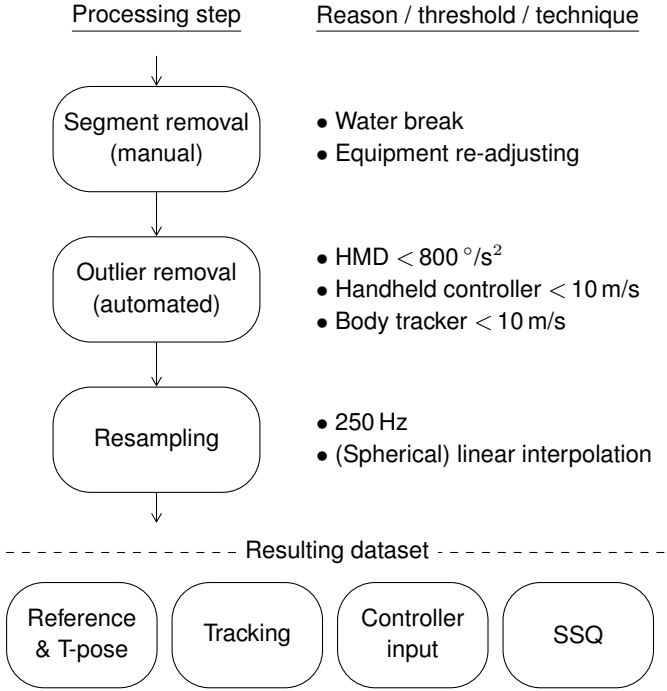
}

\narxiv{The \gls{6dof} tracking dataset is publicly\nobreakdash-accessible at~\cite{cacerumd_zenodo}. The position and orientation data are associated with a 5D grid (axes: user, application, scene, device, and time) and stored in gzip\nobreakdash-compressed HDF5 format with a total size of \SI{8}{\giga\byte}. Each sample is represented by seven data variables: three for the position $x,y,z$ (Cartesian notation) and four for the orientation $q_x, q_y, q_z, q_w$ (quaternion notation).}%
\arxiv{\subsubsection{Segment and outlier removal}
We have manually removed data segments where the user has stopped to drink water or to re\nobreakdash-adjust the \gls{xr} equipment on their body. Outliers were removed where \gls{hmd} angular acceleration is greater than \SI{800}{rad\per\second\squared} (peak value during amateur sports~\cite{miller_envelope_2020}), where handheld controller velocity surpasses \SI{10}{\meter\per\second} (martial arts punch velocity~\cite{urbinati_a_2013}), and where the body tracker's velocity exceeds \SI{10}{\meter\per\second} (athlete sprinting speed~\cite{haugen_the_2019}). Hence, the removed outliers are physically implausible, while we could not trace their origins to a particular sensor or component.\par}%
\arxiv{\subsubsection{Resampling}
The }%
\narxiv{The }published tracking data for the four devices have been synchronised in time and resampled to \SI{250}{\hertz}\arxiv{\xspace (controller and body tracker sampling frequency)}.
\arxiv{This consists of subsampling the \gls{hmd} data from \SI{250}{\hertz} and interpolating the measurement samples for all four devices to bring them onto a common time axis (equal sampling times). Linear interpolation and \gls{slerp} were used for the position and orientation, respectively. Storing data without further subsampling exceeds the sampling frequency requirements of most user movements. For example, the \gls{psd} for head orientation during an \gls{xr} museum visit recorded in \cite{blandino_head_2021} showed a drop-off of more than \SI{20}{\decibel} beyond \SI{30}{\hertz}. However, the data sampled at high frequency provide an accurate description of user mobility during fast and accelerated movements, such as running and jumping~\cite{miller_envelope_2020}. 
Moreover, the}\narxiv{The} high sampling frequency results in high\nobreakdash-resolution orientation and position tracking, enabling an accurate evaluation of \gls{mmw} link quality. For instance, during a relatively fast rotation at \SI{250}{\degree\per\second}, orientation samples are provided with a \SI{1}{\degree} accuracy.

\arxiv{\subsubsection{Resulting dataset}
The \gls{6dof} position and orientation data are associated with a 5D grid (axes: user, application, scene, device, and time) and stored in gzip\nobreakdash-compressed HDF5 format with a total size of \SI{8}{\giga\byte}. Each sample is represented by seven data variables: three for the position $x,y,z$ (Cartesian notation) and four for the orientation $q_x, q_y, q_z, q_w$ (quaternion notation). }%
We have compiled a separate dataset consisting of controller interaction (button clicks) and published it alongside the tracking dataset.\arxiv{\xspace The variables in the controller dataset correspond to individual controller inputs, while it features the same 5D grid as the tracking dataset.} Additionally, the reference and T\nobreakdash-pose measurements are accessible in~\arxiv{our database at~}\cite{cacerumd_zenodo}, together with the volunteer \glspl{ssq}.\arxiv{\xspace These can, for example, be used to assess the effects of the considered applications on user well\nobreakdash-being, which eludes the scope of the work at hand (see~\cite{dong_why_2022} for more on the topic).} Note that we began the experiment with the HTC Vive wireless adapter (\SI{60}{\giga\hertz} \gls{mmw} technology\arxiv{~\cite{vive_wireless}}) and later switched to a tethered setup due to latency and reliability issues. We noticed high \gls{ssq} scores when employing the wireless adapter, with user seven having to stop due to sickness. We changed to a cable and pulley setup (allows a high degree of user movement) starting with user 20. Hence, users 1\nobreakdash--19 were using the wireless adapter, while users 20\nobreakdash--33 were using the cable instead. This information can be valuable for assessing the effects of current \gls{mmw} technology on user sickness and to evaluate the impact of wireless \glspl{hmd} on user mobility.

\arxiv{As a final remark on the data, we have noticed that there is very little coherence among the available prior art datasets in terms of file format (e.g., CSV and MATLAB files), data types (e.g., quaternions vs. Euler axes), and coordinate conventions (e.g., the directions of the Euler axes). Our approach aims to achieve a self\nobreakdash-explanatory dataset structure, to decouple the tracking data from other inputs, and to provide coherency with the coordinate conventions of common Python packages, such as Scipy. While further evaluation in C/C++ and MATLAB would be needed, we note that the current format allows for efficient computation using tools such as Xarray, Dash, and Scipy. Nonetheless, to make the data more approachable, \cite{cacerumd_zenodo} includes tracking data in both the HDF5~format (version 1.1.0) and as individual CSV~files (version 0.1.0).}

\section{Mobility characterisation}%
\label{sec:mobility_characterisation}

This section characterises user mobility in the three considered applications and describes the results per evaluated mobility variable. With respect to the subsections, these are:
\begin{itemize}\vspace{.25em}\setlength\itemsep{.25em}
\item[A:] Planar head distance from the centre of the playspace.
\item[B:] Head vertical position, relative to the user's height.
\item[C:] Head pitch angle (looking direction up/down).
\item[D:] Lateral head velocity (combined in 3D Cartesian coord.).
\item[E:] Angular head velocity (combined yaw and pitch rate).
\item[F:] Planar distance between the hands and the head.
\item[G:] Vertical distance between the hands and the head.
\item[H:] Continuous duration that the hands spend above the head.
\item[I:] Lateral hand velocity (combined in 3D Cartesian coord.).
\end{itemize}\vspace{.25em}
\Cref{tab:mobility_characterisation} summarises the characterisation using the upper percentiles of the measured data and fitted statistical distributions. The fits are derived for multiple distributions by minimizing the negative log\nobreakdash-likelihood function. The distribution that best approximates the data is then selected according to the Kolmogorov\nobreakdash-Smirnov goodness of fit evaluation.

\subsection{Head planar position}
\label{sec:head_planar_position}

The $3\times3$~\si{\meter} playspace constrains user movement in the considered \gls{xr} applications. The rows of~\Cref{tab:mobility_characterisation} corresponding to ID A show that users stay within a single square meter, centred around the middle of the playspace, for more than \SI{50}{\percent} of the time in all three applications. 
We notice that users move the furthest in the training \gls{xr} use case (\wren), occasionally reaching the edges of the playspace. This is highlighted by the large values of the upper percentiles and the large shape and scale parameters of the Gamma distribution (long tail).

\arxiv{\Cref{fig:head_planar_position} plots the planar position distributions for the three applications. The three sets of orthogonal lines in the middle represent the principal components, while the \glspl{pdf} along the x and y direction are plotted at the top and to the right of the figure, respectively. In addition to the findings in \Cref{tab:mobility_characterisation}, we notice a covariance in the xy\nobreakdash-position in \hali and \wren, evident from the angled principal components. The covariance for these originates from car maintenance tasks and puzzles in the two applications. Users appear to have conducted these tasks more often with their right hand, which also resulted in a diagonal head and body movement. Since most volunteers were right\nobreakdash-handed, we hypothesise that the biased handedness is the reason for the position covariance. Additionally, we can notice in \Cref{fig:head_planar_position} that users in \piwi move more along the x axis then they do along the y axis. This is a consequence of the automatic in\nobreakdash-game forward movement and of the dodging of obstacles.}

\arxiv{
\begin{figure}[h]
    \centering
    \plotFig{\plotFlag}{\begin{tikzpicture}

    \begin{axis}[
        set layers=axis lines on top,
        generalPlotStyle,
        width=0.85\linewidth,
        height=0.7\linewidth,
        grid=none,
        legend style={{at={(0.02,0.02)}, anchor=south west}},
        xlabel={\small x (\SI{}{\meter})},
        x label style={at={(axis description cs:0.5,-0.1)},anchor=north},
        ylabel={\small y (\SI{}{\meter})},
        y label style={at={(axis description cs:-0.1,.5)},anchor=south},
        xmin=-1.5, xmax= 1.5,
        ymin=-1.5, ymax= 1.5,
        xtick={-1.5,-1.0,...,1.5},
        ytick={-1.5,-1.0,...,1.5},
    ]

    \addlegendimage{kulgrid2, ultra thick}
    \addlegendimage{my_red, ultra thick}
    \addlegendimage{my_pink, ultra thick}
    
    \addplot graphics [includegraphics cmd=\pgfimage, xmin=-1.5, xmax=1.5, ymin=-1.5, ymax=1.5] {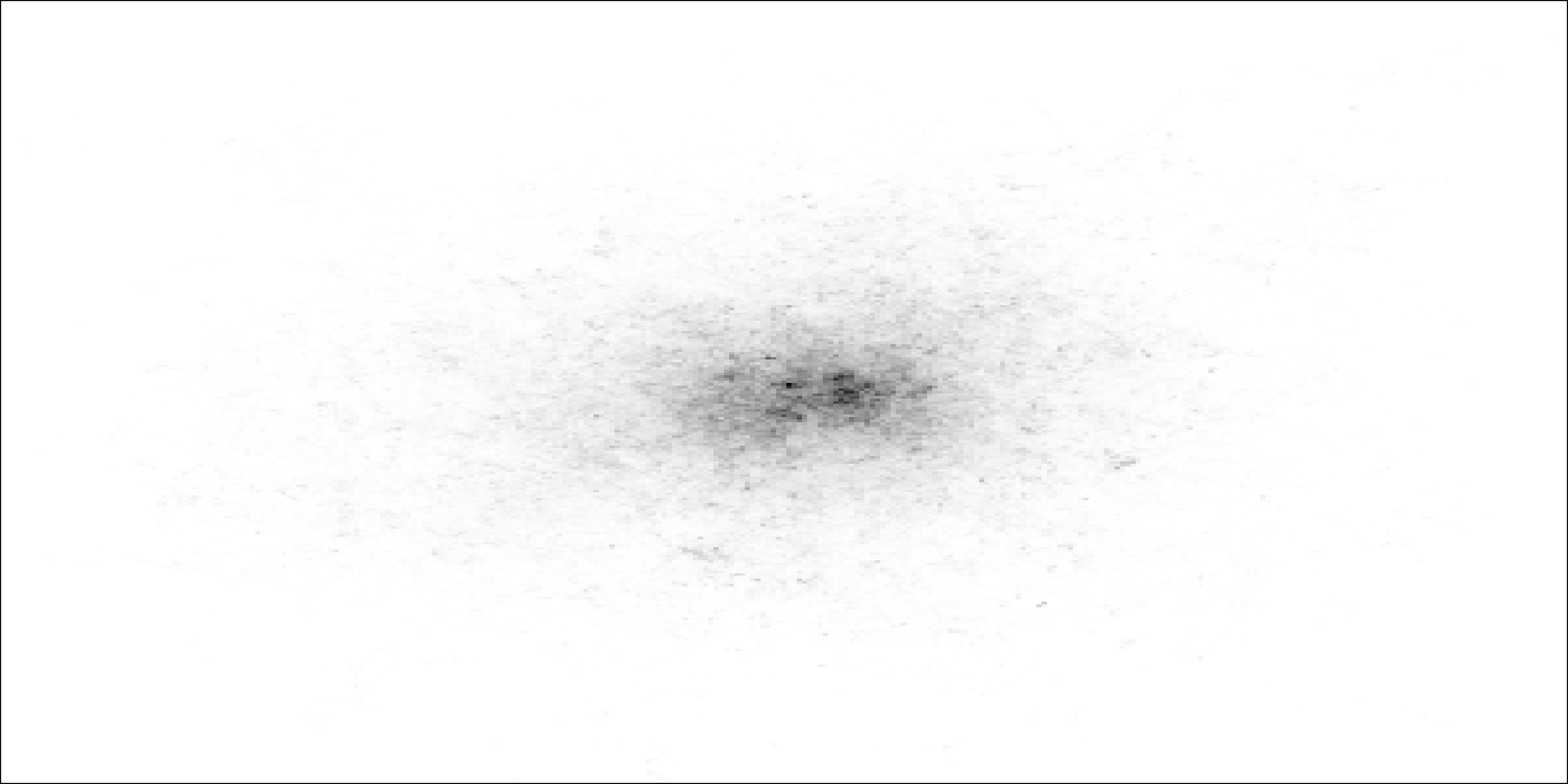};

    \draw[line width = 0.1pt, black, dashed] (axis cs: -1.5, 0) -- (axis cs: 1.5, 0);
    \draw[line width = 0.1pt, black, dashed] (axis cs: 0, -1.5) -- (axis cs: 0, 1.5);
    
    \coordinate (hl_median) at (axis cs: 0.03, -0.02);
    \draw[ultra thick, line cap=round, kulgrid2] (hl_median) -- (axis cs: 0.03 + 0.316, -0.02 + 0.179);
    \draw[ultra thick, line cap=round, kulgrid2] (hl_median) -- (axis cs: 0.03 + -0.151, -0.02 + 0.265);

    \coordinate (wr_median) at (axis cs: -0.03, 0.05);
    \draw[ultra thick, line cap=round, my_red] (wr_median) -- (axis cs: {-0.03 + sqrt(0.163)*+0.969}, {0.05 + sqrt(0.163)*+0.247});    
    \draw[ultra thick, line cap=round, my_red] (wr_median) -- (axis cs: {-0.03 + sqrt(0.144)*-0.247}, {0.05 + sqrt(0.144)*+0.969});

    \coordinate (pw_median) at (axis cs: 0.12, 0.01);
    \draw[ultra thick, line cap=round, my_pink] (pw_median) -- (axis cs: {0.12 + sqrt(0.122)*+0.999}, {0.01 + sqrt(0.122)*+0.039});
    \draw[ultra thick, line cap=round, my_pink] (pw_median) -- (axis cs: {0.12 + sqrt(0.042)*-0.039}, {0.01 + sqrt(0.042)*+0.999});

    \legend{Half Life: Alyx, Wrench, Pistol Whip};
        
    \end{axis}

    \begin{axis}[
        set layers=axis lines on top,
        at={(0,0.548\linewidth)},
        generalPlotStyle,
        width=0.85\linewidth,
        height=0.3\linewidth,
        xmin=-1.5, xmax= 1.5,
        grid=none,
        ylabel={\small PDF (x)},
        y label style={at={(axis description cs:-0.1,.5)},anchor=south},
        xticklabels={},
        yticklabels={},
        scaled y ticks=false
    ]

    \addplot [kulgrid2, very thick] table[x=abscissa, y=hl] {\headPositionPdfX};
    \addplot [my_red, very thick] table[x=abscissa, y=wr] {\headPositionPdfX};
    \addplot [my_pink, very thick] table[x=abscissa, y=pw] {\headPositionPdfX};
        
    \end{axis}

    \begin{axis}[
        set layers=axis lines on top,
        at={(0.7\linewidth,0)},
        generalPlotStyle,
        width=0.3\linewidth,
        height=0.7\linewidth,
        ymin=-1.5, ymax= 1.5,
        grid=none,
        xlabel={\small PDF (y)},
        x label style={at={(axis description cs:0.5,-0.1)},anchor=north},
        xticklabels={},
        yticklabels={},
        scaled x ticks=false
    ]

    \addplot [kulgrid2, very thick] table[y=abscissa, x=hl] {\headPositionPdfY};
    \addplot [my_red, very thick] table[y=abscissa, x=wr] {\headPositionPdfY};
    \addplot [my_pink, very thick] table[y=abscissa, x=pw] {\headPositionPdfY};
        
    \end{axis}

\end{tikzpicture}}
    \caption{Planar user position within the playspace. The principal components per application are plotted on top of the position \gls{pdf} of Half Life: Alyx (a higher probability is darker). The \gls{pdf} for the x and y axis is shown at the top and right, respectively. Dashed black lines represent zero-values.}
    \label{fig:head_planar_position}
\end{figure}
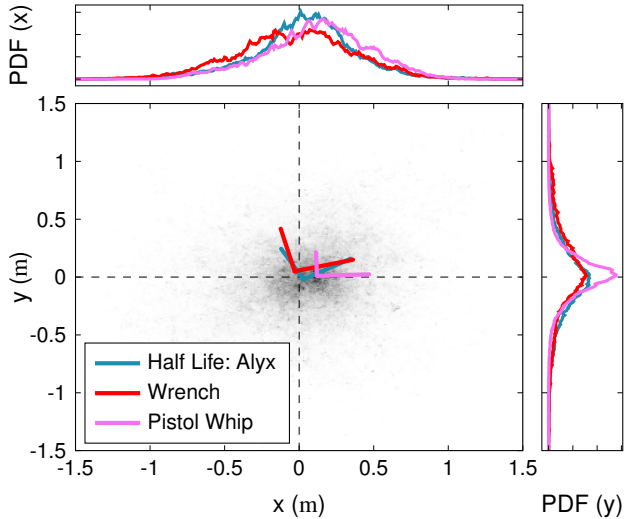
}

\subsection{Head vertical position}
\label{sec:head_vertical_position}

\Cref{tab:mobility_characterisation}, ID B, lists the ratio between the \glsposs{hmd} vertical position when the user is standing upright, recorded during the preliminary T\nobreakdash-pose measurement, and during gameplay. This vertical displacement is presented on a relative scale, since user height can vary considerably\arxiv{, making a comparison in absolute terms less insightful}. We notice that users crouch down more in \gls{xr} training and less in \gls{xr} games, shown by the 2\nobreakdash-times larger scale parameter of the Cauchy distribution for \wren (medium gray, middle row) in comparison to \hali and \piwi. The crouching in \wren appeared mostly when the users were working on a car part close to the floor or even lying underneath the car.\arxiv{\xspace On the other hand, \hali and \piwi require less pronounced crouching.}

\arxiv{
\Cref{fig:head_vertical_factor} plots the \gls{pdf} of the head's vertical position to highlight the differences between the applications. Namely, showing the wider distribution of \wren. Notice that a vertical position factor greater than one was also recorded since users would displace the \glsposs{hmd} body frame (position) by up to \SI{10}{\centi\meter} when looking up.
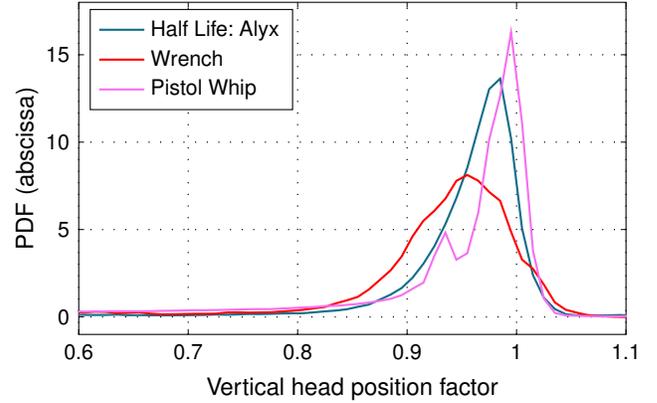
\begin{figure}[h]
    \centering
    \plotFig{\plotFlag}{\begin{tikzpicture}[
    spy using outlines={black, rounded rectangle, width=3.8cm, height=1cm, connect spies}
]

    \begin{axis}[
        generalPlotStyle,
        width=\linewidth,
        legend style={
            at={(0.02,0.98)},
            anchor=north west
        },
        xlabel={\small Vertical head position factor},
        ylabel={\small PDF (abscissa)},
        xmin=0.6,
        xmax=1.1,
        ymin= -1,
        ymax=18,
    ]
    
    \draw[very thin, black!70] (axis cs: -0.08, 97.5) -- (axis cs: 0.08, 97.5) node [anchor=east, black, xshift=-0.8cm] {\fontsize{4}{6}\selectfont 97.5};
    \draw[very thin, black!70] (axis cs: -0.08, 95) -- (axis cs: 0.08, 95) node [anchor=west, black, xshift=-0.00cm] {\fontsize{4}{6}\selectfont 95};

    \addplot [kulgrid3, thick] table[x=abscissa, y=hl] {\headVerticalFactorPdf};
    \addplot [my_red, thick] table[x=abscissa, y=wr] {\headVerticalFactorPdf};
    \addplot [my_pink, thick] table[x=abscissa, y=pw] {\headVerticalFactorPdf};
    

    \legend{Half Life: Alyx, Wrench, Pistol Whip};
    
  \end{axis}
\end{tikzpicture}}
    \caption{\Gls{pdf} of the head's vertical position. Values less than one mean the user's head was at a lower height than during the T-pose measurement.}
    \label{fig:head_vertical_factor}
\end{figure}
}

\subsection{Head orientation}
\label{sec:head_orientation}

\begin{figure}[h]
    \centering
    \plotFig{\plotFlag}{\begin{tikzpicture}

    \begin{axis}[
        set layers=axis lines on top,
        generalPlotStyle,
        width=0.85\linewidth,
        height=0.7\linewidth,
        grid=none,
        legend style={{at={(0.02,0.02)}, anchor=south west}},
        xlabel={\small Yaw (\SI{}{\degree})},
        x label style={at={(axis description cs:0.5,-0.1)},anchor=north},
        ylabel={\small Pitch (\SI{}{\degree})},
        y label style={at={(axis description cs:-0.1,.5)},anchor=south},
        xmin=-180, xmax= 180,
        ymin=-90, ymax= 90,
        xtick={-180,-135,...,180},
        ytick={-90,-45,...,90},
    ]

    \addlegendimage{kulgrid2, ultra thick}
    \addlegendimage{my_red, ultra thick}
    \addlegendimage{my_pink, ultra thick}
    
    \addplot graphics [includegraphics cmd=\pgfimage, xmin=-180, xmax=180, ymin=-90, ymax=90] {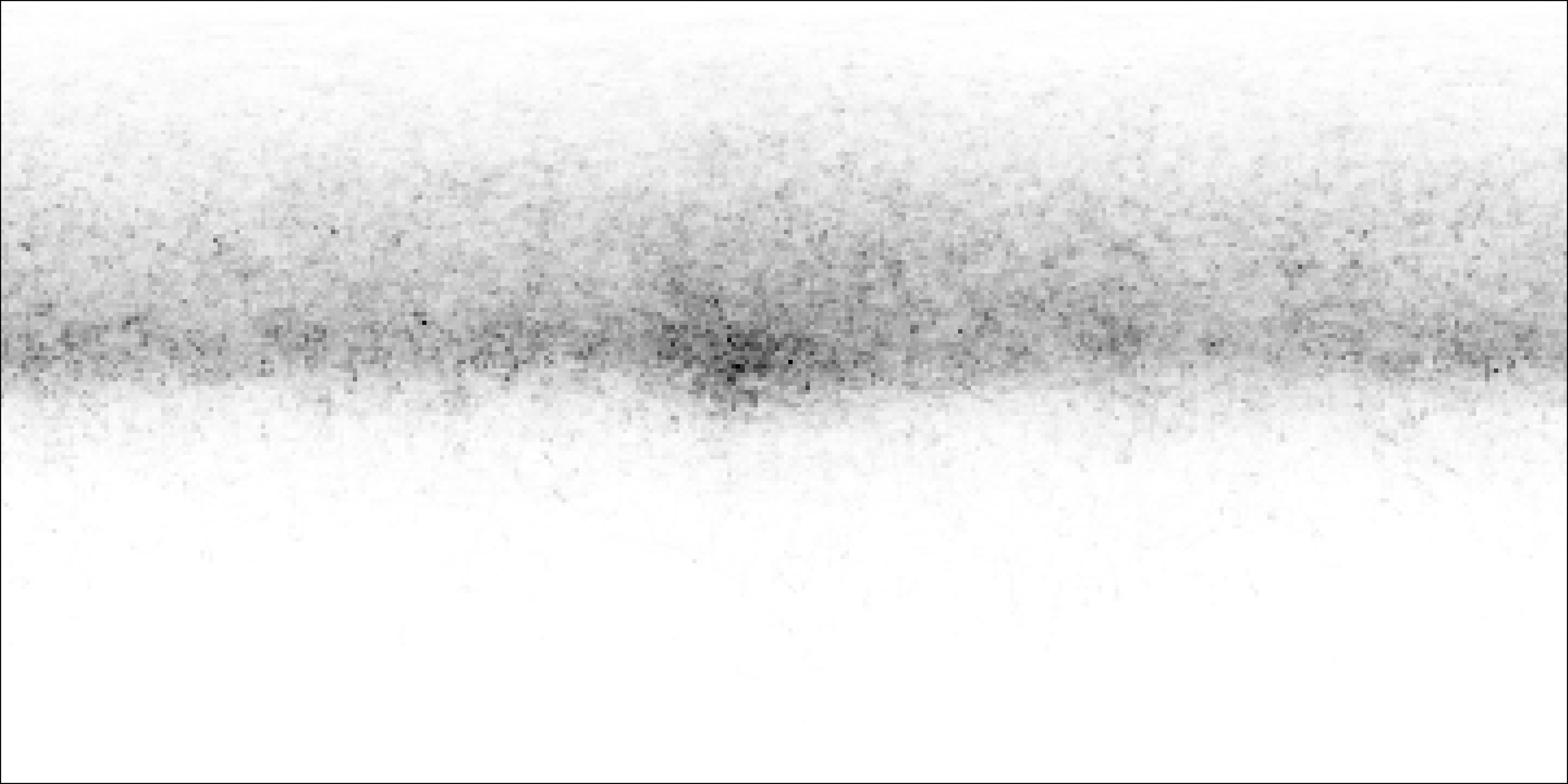};

    \draw[line width = 0.1pt, black, dashed] (axis cs: -180, 0) -- (axis cs: 180, 0);
    \draw[line width = 0.1pt, black, dashed] (axis cs: 0, -90) -- (axis cs: 0, 90);
    
    \coordinate (hl_median) at (axis cs: -1, 23);
    \draw[ultra thick, kulgrid2] (hl_median) -- (axis cs: {-1 + sqrt(9614.247)*0.999}, {23 + sqrt(9614.247)*0.001});
    \draw[ultra thick, kulgrid2] (hl_median) -- (axis cs: {-1 + sqrt(363.553)*-0.001}, {23 + sqrt(363.553)*0.999});

    \coordinate (wr_median) at (axis cs: -10, 28);
    \draw[ultra thick, my_red] (wr_median) -- (axis cs: {-10 + sqrt(6137.748)*0.999}, {28 + sqrt(6137.748)*0.001});    
    \draw[ultra thick, my_red] (wr_median) -- (axis cs: {-10 + sqrt(710.162)*-0.001}, {28 + sqrt(710.162)*0.999});

    \coordinate (pw_median) at (axis cs: 0, 3);
    \draw[ultra thick, my_pink] (pw_median) -- (axis cs: {0 + sqrt(544.763)*0.999}, {3 + sqrt(544.763)*0.048});
    \draw[ultra thick, my_pink] (pw_median) -- (axis cs: {0 + sqrt(105.187)*-0.048}, {3 + sqrt(105.187)*0.999});

    \legend{Half Life: Alyx, Wrench, Pistol Whip};
        
    \end{axis}

    \begin{axis}[
        set layers=axis lines on top,
        at={(0,0.548\linewidth)},
        generalPlotStyle,
        width=0.85\linewidth,
        height=0.3\linewidth,
        xmin=-180, xmax= 180,
        grid=none,
        ylabel={\small PDF (yaw)},
        y label style={at={(axis description cs:-0.1,.5)},anchor=south},
        xticklabels={},
        yticklabels={},
        scaled y ticks=false
    ]

    \addplot [kulgrid2, very thick] table[x=abscissa, y=hl] {\headOrientationPdfYaw};
    \addplot [my_red, very thick] table[x=abscissa, y=wr] {\headOrientationPdfYaw};
    \addplot [my_pink, very thick] table[x=abscissa, y=pw] {\headOrientationPdfYaw};
        
    \end{axis}

    \begin{axis}[
        set layers=axis lines on top,
        at={(0.7\linewidth,0)},
        generalPlotStyle,
        width=0.3\linewidth,
        height=0.7\linewidth,
        ymin=-90, ymax= 90,
        grid=none,
        xlabel={\small PDF (pitch)},
        x label style={at={(axis description cs:0.5,-0.1)},anchor=north},
        xticklabels={},
        yticklabels={},
        scaled x ticks=false
    ]

    \addplot [kulgrid2, very thick] table[y=abscissa, x=hl] {\headOrientationPdfPitch};
    \addplot [my_red, very thick] table[y=abscissa, x=wr] {\headOrientationPdfPitch};
    \addplot [my_pink, very thick] table[y=abscissa, x=pw] {\headOrientationPdfPitch};
        
    \end{axis}

\end{tikzpicture}}
    \caption{Head orientation principal components per application on top of the orientation \gls{pdf} of \hali (darker corresponds to a higher probability). The \glspl{pdf} for yaw and pitch are shown at the top and right, respectively. Dashed black lines represent zero-yaw and zero-pitch. Recall that a positive yaw is caused by leftward head rotation, and a positive pitch means the user is looking below the horizon.}
    \label{fig:head_orientation_heatmap}
\end{figure}
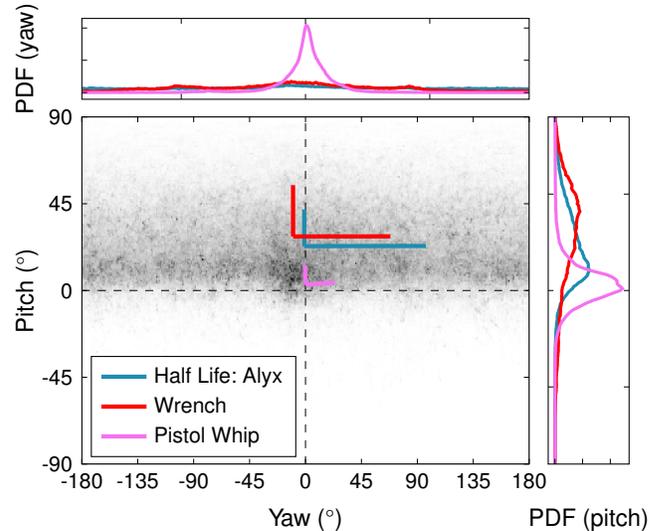

\Cref{fig:head_orientation_heatmap} shows the two principal components of user head orientation in the yaw\nobreakdash-pitch domain. We notice that all applications feature a positive pitch bias (gaze below the horizon). A bias of up to \SI{10}{\degree} has already been reported in prior art~\cite{corbillon_360-degree_2017, blandino_head_2021}, however, the median pitch listed in \Cref{tab:mobility_characterisation}, ID C, for \hali and \wren is considerably larger at \SI{20}{\degree} and \SI{32}{\degree}, respectively. \arxiv{Note that the principal components in \Cref{fig:head_orientation_heatmap} start from the mean orientation, which is \SI{23}{\degree} and \SI{28}{\degree}, correspondingly.} The tendency to look down (positive pitch)\arxiv{\xspace in \hali and \wren} is also visible from the \glspl{pdf} of the pitch, plotted at the right of \Cref{fig:head_orientation_heatmap}, best approximated using skewed normal distributions. Conversely, the symmetrical shape and heavy tails make \piwi conform with the logistic distribution. Note that the \SI{90}{\degree} pitch in \Cref{tab:mobility_characterisation} occurred due to users locating themselves relative to a floor marking at the centre of the playspace.

In terms of yaw, users in \hali and \wren explored all orientations, \narxiv{while }%
\arxiv{with a slight bias in the forward\nobreakdash-facing direction. This is because objects are placed in front of the user when starting the application. For example, the car in \wren is placed at a yaw of approximately \SI{0}{\degree}. Moreover, }%
we can observe a negative yaw bias for \wren, originating from predominant right\nobreakdash-hand usage. For example, when reaching for a part under the car's bonnet, the users looked towards their right hand. Since most volunteers were right\nobreakdash-handed, we hypothesise that the biased handedness caused the orientation bias.

\subsection{Head lateral velocity}
\label{sec:head_lateral_velocity}

\Cref{tab:mobility_characterisation}, ID D, shows the lateral head velocity in each application. That is, the 3\nobreakdash-dimensional velocity in a Cartesian coordinate system. We see that users exhibit the fastest lateral movement in \piwi, while the largest value across all applications is \SI{3.2}{\meter\per\second}. Lateral velocity is roughly exponentially distributed, while the Weibull distribution further improves the goodness of fit for the upper percentiles due to its heavier tail. The measured lateral velocity is well\nobreakdash-below the velocities observed in other domains, such as vehicular communications or even general sports activities. Instead, \gls{xr} use cases are characterised by fast rotations.

\subsection{Head angular velocity}
\label{sec:head_angular_velocity}

\Cref{tab:mobility_characterisation}, ID E, lists the combined angular velocity along the pitch and yaw axis, which would be observed by a forward\nobreakdash-facing antenna. We notice that users can rotate their heads in excess of \SI{360}{\degree\per\second}, as indicated in prior art~\cite{lincoln_low_2017}. 
However, for the majority of time, the angular velocity is contained below \SI{100}{\degree\per\second} in all three applications. Head angular velocity is best described using the log\nobreakdash-normal distribution, since the angular velocity's logarithm is approximately normally distributed. This is also the mobility variable with the best fitting statistical distribution.
\arxiv{\Cref{fig:head_angular_velocity_percentiles} confirms that \wren features the lowest head angular velocity of the three applications. Users exhibit similar angular velocity in \hali and \piwi, with the former featuring higher peak values. Users in \piwi exhibit higher angular head velocity at lower percentiles, while the highlighted intersection in \Cref{fig:head_angular_velocity_percentiles} indicates the longer tail of \hali' distribution.}
In view of the later evaluation of the possible impact on \gls{mmw} links, we are interested in understanding whether the high velocity occurs randomly or in bursts.

\arxiv{
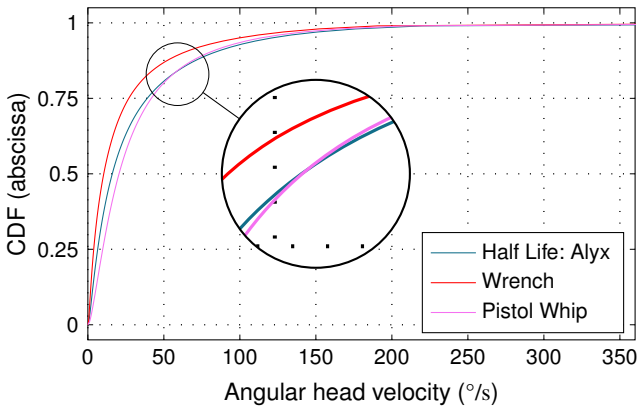
\begin{figure}[h]
    \centering
    \plotFig{\plotFlag}{\begin{tikzpicture}[
    spy using outlines={black, rounded rectangle, width=2.5cm, height=2.5cm, connect spies}
]

    \begin{axis}[
        generalPlotStyle,
        width=\linewidth,
        legend style={
            at={(0.98,0.02)},
            anchor=south east
        },
        xlabel={\small Angular head velocity (\SI{}{\degree\per\second})},
        ylabel={\small CDF (abscissa)},
        xmin=0,
        xmax=360,
        ymin= -0.05,
        ymax=1.05,
        ytick={0, 0.25, 0.50, 0.75, 1.0},
    ]
    
    \draw[very thin, black!70] (axis cs: -0.08, 97.5) -- (axis cs: 0.08, 97.5) node [anchor=east, black, xshift=-0.8cm] {\fontsize{4}{6}\selectfont 97.5};
    \draw[very thin, black!70] (axis cs: -0.08, 95) -- (axis cs: 0.08, 95) node [anchor=west, black, xshift=-0.00cm] {\fontsize{4}{6}\selectfont 95};

    \addplot [kulgrid3] table[y=percentiles, x=hl] {\headAngVelPerc};
    \addplot [my_red] table[y=percentiles, x=wr] {\headAngVelPerc};
    \addplot [my_pink] table[y=percentiles, x=pw] {\headAngVelPerc};
        
    \coordinate (spypoint) at (axis cs: 59,0.83);
    \coordinate (magnifyglass) at (axis cs: 150, 0.5);
    \spy [fill=red, magnification=3.0] on (spypoint) in node [fill=white] at (magnifyglass);

    \legend{Half Life: Alyx, Wrench, Pistol Whip};
    
  \end{axis}
\end{tikzpicture}}
    \caption{\Gls{cdf} of the head angular velocity.}
    \label{fig:head_angular_velocity_percentiles}
\end{figure}
}

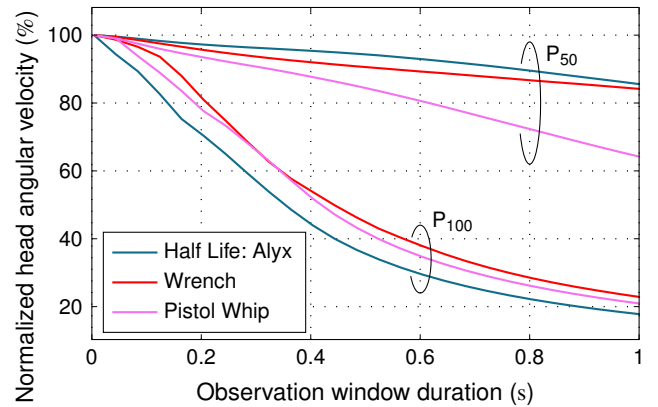
\begin{figure}[h]
    \centering
    \plotFig{\plotFlag}{\begin{tikzpicture}[]

    \begin{axis}[
        set layers=axis lines on top,
        generalPlotStyle,
        width=\linewidth,
        xlabel={\small Observation window duration (\SI{}{\second})},
        ylabel={\small Normalized head angular velocity (\SI{}{\percent})},
        y label style={at={(axis description cs:-0.1,.4)},anchor=south},
        xmin=0,
        xmax=1,
        legend style={{at={(0.02,0.02)}, anchor=south west}},
    ]

    \addplot [kulgrid3, thick] table[x=window_size, y=hl_100] {\meanAngularVelocity};
    \addplot [my_red, thick] table[x=window_size, y=wr_100] {\meanAngularVelocity};
    \addplot [my_pink, thick] table[x=window_size, y=pw_100] {\meanAngularVelocity};
    
       
    \addplot [kulgrid3, thick] table[x=window_size, y=hl_050] {\meanAngularVelocity};
    \addplot [my_red, thick] table[x=window_size, y=wr_050] {\meanAngularVelocity};
    \addplot [my_pink, thick] table[x=window_size, y=pw_050] {\meanAngularVelocity};
    
    
    
    
    \node[black, anchor=center] (p50) at (axis cs: 0.86, 94) {P$_{\text{50}}$}; 
    \draw[black] (0.8, 86) [partial ellipse=-150:140:0.02 and 18];
    
    \node[black, anchor=center] (p100) at (axis cs: 0.66, 45) {P$_{\text{100}}$};
    \draw[black] (0.6, 40) [partial ellipse=-150:140:0.02 and 10];

    \legend{Half Life: Alyx, Wrench, Pistol Whip};

  \end{axis}
\end{tikzpicture}}
    \caption{Running average of the head's angular velocity, dependent on observation window length and normalized relative to the velocity for a \SI{4}{\milli\second} window. Median and maximal velocity are marked by P$_{\text{50}}$ and P$_{\text{100}}$, correspondingly.}
    \label{fig:head_angular_velocity_vs_window_size}
\end{figure}

\Cref{fig:head_angular_velocity_vs_window_size} shows that the running average of the head's angular velocity during a \SI{0.1}{\second} observation window retains \SI{90}{\percent} of its short\nobreakdash-term value. 
\arxiv{
Thus, multiplying the angular velocity values in \Cref{tab:mobility_characterisation}, ID E, by the factor $0.90$ and further by the window size of \SI{0.1}{\second} allows us to calculate the head rotation during a \SI{0.1}{\second} window. }%
Using \hali as an example, we can estimate the P$_{\text{50}}$ and P$_{\text{100}}$ rotation over the course of \SI{0.1}{\second} at approximately \SI{1.5}{\degree} and \SI{60}{\degree}, respectively.%
\arxiv{\xspace The same principle can be applied for other window sizes and applications, where it is evident that the decrease in angular velocity noticeably diverges for longer windows. For instance, the maximal recorded running average of the angular velocity within a window size of \SI{1}{\second}, reduces to about \SI{20}{\percent} of the maximal short\nobreakdash-term velocity (\SI{4}{\milli\second} window size). Conversely, the running average of the median velocity reduces to \SIrange{60}{80}{\percent} of its short\nobreakdash-term value.}
We notice that the median of the angular velocity's running average reduces the most with in \piwi as window size increases, indicating predominantly shorter\nobreakdash-term rotations. A similar trend was observed for other percentiles, which are not plotted in view of clarity. The maximum of the angular velocity's running average rapidly decreases with the window size, reaching less than \SI{360}{\degree\per\second} beyond a window size of \SI{0.3}{\s} for all three applications. Note that the 2D angular velocity can be up to a factor of $\sqrt{2}$ larger than the velocity around a single axis of rotation. 

\subsection{Planar hand-to-head distance}
\label{sec:planar_hand_position}

\Cref{tab:mobility_characterisation}, ID F, shows that users mostly kept their hands close to their body, indicated by the \SIrange{0.2}{0.3}{\meter} median (P$_{50}$) planar distance (parallel to the floor) between the user's hands and head. The distance's \gls{pdf} has a sharp cut\nobreakdash-in (steep increase at short distances) and, according to P$_{98}$--P$_{100}$, a long tail. Given the distance's logarithm is normally distributed and that the planar positions of the hands and \gls{hmd} rarely coincide, the measurements are best fitted using a log-normal distribution with the shape parameter set to less than one (clear peak).
\arxiv{
\Cref{fig:hand_planar_distance} shows the \glspl{pdf} of the planar hand distance from the \gls{hmd}. 
\begin{figure}[h]
    \centering
    \plotFig{\plotFlag}{\begin{tikzpicture}[
    spy using outlines={black, rounded rectangle, width=3.8cm, height=1cm, connect spies}
]

    \begin{axis}[
        generalPlotStyle,
        width=\linewidth,
        legend style={
            at={(0.98,0.98)},
            anchor=north east
        },
        xlabel={\small Planar hand distance to the HMD (m)},
        ylabel={\small PDF (abscissa)},
        xmin=0,
        xmax=1.2,
        ymin= -0.2,
        ymax=5.6,
    ]
    
    \draw[very thin, black!70] (axis cs: -0.08, 97.5) -- (axis cs: 0.08, 97.5) node [anchor=east, black, xshift=-0.8cm] {\fontsize{4}{6}\selectfont 97.5};
    \draw[very thin, black!70] (axis cs: -0.08, 95) -- (axis cs: 0.08, 95) node [anchor=west, black, xshift=-0.00cm] {\fontsize{4}{6}\selectfont 95};

    \addplot [kulgrid3] table[x=abscissa, y=hl] {\handPlanarDistancePdf};
    \addplot [my_red] table[x=abscissa, y=wr] {\handPlanarDistancePdf};
    \addplot [my_pink] table[x=abscissa, y=pw] {\handPlanarDistancePdf};

    \legend{Half Life: Alyx, Wrench, Pistol Whip};
    
  \end{axis}
\end{tikzpicture}}
    \caption{\Gls{pdf} of the planar distance between the user's hands and the \gls{hmd}.}
    \label{fig:hand_planar_distance}
\end{figure}
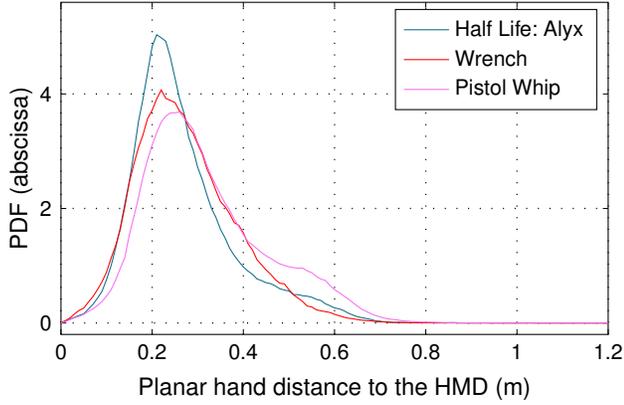
}

\subsection{Vertical hand-to-head distance}
\label{sec:vertical_hand_position}

\Cref{tab:mobility_characterisation}, ID G, lists the vertical distance between the users' hands and the \gls{hmd}. In \wren, users tend to use their hands in a narrow region with almost equal likelihood of raising or lowering them, represented by the logistic distribution. \hali and \piwi are better approximated by the longer tails of a Weibull distribution since users often lowered their hands when they were not in a shootout. We notice that, in all three applications, the hands spend roughly \SI{3}{\percent} of time at or above head height.\arxiv{\xspace Moreover, considering a \SIrange{10}{15}{\centi\meter} vertical distance between the \gls{hmd} (eyes) and the user's chin, we see that users keep their hands below the head roughly \SI{90}{\percent} of the time.} Similar to the head's angular velocity, this begs the question: 
do users place their hands above the head for a long duration, or does the movement resemble a quicker motion, similar to waving?

\arxiv{
\Cref{fig:hand_vertical_distance} shows the vertical position of the hands relative to the \gls{hmd}. The main difference between the applications in terms of vertical hand distribution is that users tend to use their hands in \piwi at a lower height, compared to \hali and Wrench. This is because of the game's design, where players have to often lower their hands in order to reload their firearm.

\begin{figure}[h]
    \centering
    \plotFig{\plotFlag}{\begin{tikzpicture}[
    spy using outlines={black, rounded rectangle, width=3.8cm, height=1cm, connect spies}
]

    \begin{axis}[
        generalPlotStyle,
        width=\linewidth,
        xlabel={\small Vertical hand position relative to the HMD (m)},
        ylabel={\small CDF (\%)},
        xmin=-1,
        xmax=0.4,
        ymin= -0.05,
        ymax=1.05,
        ytick={0, 0.25, 0.50, 0.75, 1},
        yticklabels={0, 0.25, 0.50, 0.75, 1}
    ]
    
    \draw[very thin, black!70] (axis cs: -0.08, 0.975) -- (axis cs: 0.08, 0.975) node [anchor=east, black, xshift=-0.8cm] {\fontsize{4}{6}\selectfont 0.975};
    \draw[very thin, black!70] (axis cs: -0.08, 0.95) -- (axis cs: 0.08, 0.95) node [anchor=west, black, xshift=-0.00cm] {\fontsize{4}{6}\selectfont 0.95};

    \addplot [kulgrid3] table[x=hl, y=percentiles] {\handVerticalDistancePercentiles};
    \addplot [my_red] table[x=wr, y=percentiles] {\handVerticalDistancePercentiles};
    \addplot [my_pink] table[x=pw, y=percentiles] {\handVerticalDistancePercentiles};
    
    \coordinate (spypoint) at (axis cs: 0,0.96);
    \coordinate (magnifyglass) at (axis cs: 0, 0.25);
    \spy [fill=red, magnification=2.0] on (spypoint) in node [fill=white] at (magnifyglass);

    \legend{Half Life: Alyx, Wrench, Pistol Whip};
    
  \end{axis}
\end{tikzpicture}}
    \caption{\Gls{cdf} of the vertical distance between the user's head and hands.}
    \label{fig:hand_vertical_distance}
\end{figure}
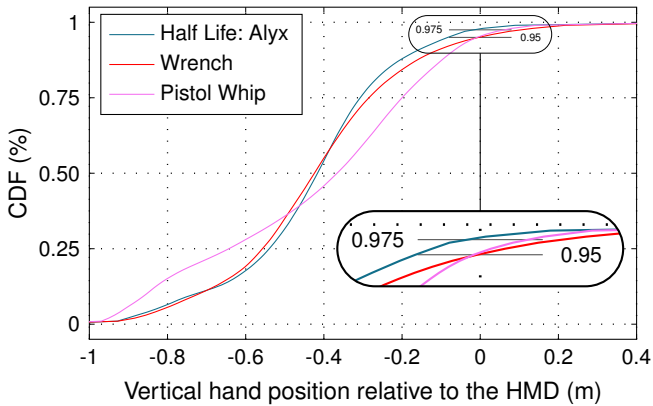
}

\subsection{Hand persistence above the head}
\label{sec:hand_persistence}

\Cref{tab:mobility_characterisation}, ID H, shows that users raise their hands above the \gls{hmd} and lower them back down within the first \SIrange{0.4}{1.3}{\second} on half of the occasions (P$_{50}$). Furthermore, this duration can grow to more than \SI{20}{\second} in \gls{xr} games (\hali and \piwi) and even up to \SI{45}{\second} during the considered \gls{xr} training (\wren) -- an important finding for \gls{mmw} systems, that has not been reported before.\arxiv{\xspace In general, the longest hand persistence above the head was recorded in \wren, where users also carried out maintenance tasks while the car was lifted above them. \piwi showed the least hand persistence, since users would only quickly aim their firearm and then lower their hand. With a mixture of puzzles and action sequences, \hali featured moderate hand usage above the head.} The slow \gls{pdf} decay can be well represented using the long tails of a Weibull distribution with the shape parameter set to less than one.

\arxiv{
\Cref{fig:hand_persistence} plots the \glspl{pdf} of the hand persistence to highlight the differences between the applications. We notice how \wren features a wider distribution with a larger positive offset, compared to \hali and \piwi. The latter two have similar \glspl{pdf}, with \hali exhibiting somewhat longer hand usage above the head.
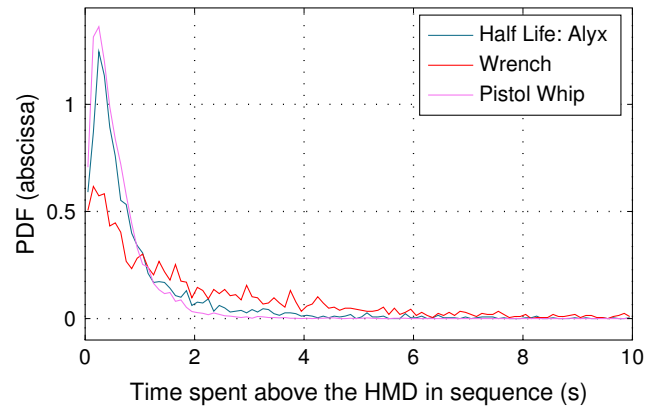
\begin{figure}[h]
    \centering
    \plotFig{\plotFlag}{\begin{tikzpicture}[
    spy using outlines={black, rounded rectangle, width=3.8cm, height=1cm, connect spies}
]

    \begin{axis}[
        set layers=axis lines on top,
        generalPlotStyle,
        width=\linewidth,
        xlabel={\small Time spent above the HMD in sequence (s)},
        ylabel={\small PDF (abscissa)},
        xmin=0,
        xmax=10,
        legend style={{at={(0.98,0.98)}, anchor=north east}},
        ymin= -0.1,
        ymax=1.45,
    ]
    
    \addplot [kulgrid3] table[x=abscissa, y=hl] {\handPersistencePdf};
    \addplot [my_red] table[x=abscissa, y=wr] {\handPersistencePdf};
    \addplot [my_pink] table[x=abscissa, y=pw] {\handPersistencePdf};

    \legend{Half Life: Alyx, Wrench, Pistol Whip};
    
  \end{axis}
\end{tikzpicture}}
    \caption{\Glspl{pdf} showing the likelihood of the user's hands staying above the \gls{hmd} for a given duration of time.}
    \label{fig:hand_persistence}
\end{figure}
}

\subsection{Hand velocity}

\Cref{tab:mobility_characterisation}, ID I, shows that hand velocity is the highest in \piwi, which requires users to punch and aim at moving targets.\arxiv{\xspace Users exhibit roughly \SI{65}{\percent} lower hand velocity in \hali and \wren, where there are no tasks which would require fast hand movement.} We notice a high maximum hand velocity in all applications, however, since \SI{10}{\meter\per\second} is the punching velocity of martial artists, this could also indicate that the P$_{100}$ values are outliers~\cite{urbinati_a_2013}. For example, due to an underperforming tracking system on the handheld controllers, which are often obstructed by the user and feature fewer \gls{ir} photodiodes than the \gls{hmd}.
\arxiv{Concluding on the large P$_{100}$ values would require further evaluation of the \gls{ir} tracking system's accuracy during controller movement (\cite{niehorster_accuracy_2017} provides a comprehensive assessment for a static \gls{hmd}).} Hand lateral velocity shows a similar distribution to \gls{hmd} velocity, best approximated using a Weibull distribution.

\section{Impact of mobility on mmWave links}
\label{sec:impact_on_mmwave}

\begin{table*}[t]
    \centering    
    \renewcommand{\arraystretch}{1.25}
    \caption{Observations regarding the mobility characterisation and the possible impact on \gls{mmw} links.}
    \label{tab:mobility_effects_on_mmwave}
\begin{tabular}{p{0.15\textwidth}p{0.28\textwidth}p{0.32\textwidth}p{0.15\textwidth}}

\toprule
\textbf{Mobility} & \textbf{Observation} & \textbf{Possible impact on \gls{mmw} links} & \textbf{Adversity} \\ 
\midrule
    
\rowcolor{pw_gray} Head planar position & Predominantly static. & ~ & \\ 

\rowcolor{pw_gray} Head vertical position & Can crouch down to \SI{20}{\percent} of body height. & \multirow{-2}{0.32\textwidth}{Users can obstruct each other's LoS, causing attenuation of up to \SI{15}{\decibel}.} & \multirow{-2}{0.15\textwidth}{External-shadowing} \\ 

\rowcolor{wr_gray} Hand planar position & Distance to HMD is about \SIrange{20}{30}{\centi\meter}. & ~ & \\ 

\rowcolor{wr_gray} Hand vertical position & Hands are above the head for \SI{3}{\percent} of time. & ~ & \\

\rowcolor{wr_gray} Hand persistence & Can remain above head for up to \SI{45}{\second}. & \multirow{-3}{0.32\textwidth}{Hands can obstruct a large proportion of the \glsposs{hmd} view, attenuating the \acrshort{los} by several \SI{}{\decibel}.} & \multirow{-3}{0.15\textwidth}{Self-shadowing} \\ 

\rowcolor{pw_gray} Head yaw orientation & Can rotate for a full \SI{360}{\degree}. & ~ & \\ 

\rowcolor{pw_gray} Head pitch orientation & Bias of up to \SI{32}{\degree} (gaze below the horizon). & \multirow{-2}{0.32\textwidth}{Losses of several \SI{}{\decibel} or more if insufficient azimuth and elevation coverage are provided.} &  \multirow{-2}{0.15\textwidth}{Antenna misalignment} \\ 

\rowcolor{wr_gray} Head lateral velocity & Slow lateral movement of up to \SI{3}{\meter\per\second}. & ~ & \\ 
\rowcolor{wr_gray} Head angular velocity & Fast rotation, possibly exceeding \SI{360}{\degree\per\second}. & \multirow{-2}{0.32\textwidth}{Losses of several \SI{}{\decibel} or more, if beam steering at the \gls{hmd} is not applied frequently enough.} & \multirow{-2}{0.15\textwidth}{Beam misalignment} \\ 

\bottomrule
    
\end{tabular}
\end{table*}

In assessing how \gls{hmd} mobility negatively impacts \gls{mmw} links, two groups of adversities prevail. The first is \emph{shadowing}, where the \gls{mmw} link becomes attenuated due to a person or object intercepting it. The second is \emph{misalignment}, where one or more \gls{hmd} antennas (or arrays) have been moved or rotated by an extent that degrades the \gls{mmw} link. The following two sections describe the two adversity groups with regard to general \gls{hmd} mobility, while \Cref{tab:mobility_effects_on_mmwave} summarizes the observations from \Cref{sec:mobility_characterisation} and lists the possible impact on the \gls{mmw} link.

\subsection{External and self-shadowing}
\label{sec:impact_on_mmwave:shadowing}

Shadowing occurs when a person or an object obstructs a high\nobreakdash-gain propagation path between the \gls{hmd} and \gls{ap}. This is typically the \gls{los} component, where a high degree of power is concentrated in \gls{mmw} channels~\cite{cai_dynamic_2020}. Under the assumption that \glspl{ap} are deployed in a way where there are no objects obstructing the \gls{los} across the entire playspace, there are still two major possible sources of shadowing in \gls{xr} use cases: other persons 
and one's own hands~\cite{abari_enabling_2017, chukhno_interplay_2022}. We name the two \emph{external}~and \emph{self\nobreakdash-shadowing}, respectively. 

\emph{External shadowing} arises when a third person crosses the \gls{los}, usually obstructing it with the torso, neck, or head. Prior art \cite{gustafson_characterization_2012} has shown that the resulting \gls{mmw} link attenuation can be approximated analytically, for example, using the \gls{gtd}, which depends on the shape of the body part and its distance to the \gls{hmd} and the \gls{ap}. 
According to the \gls{gtd} and \gls{mmw} channel measurements, the obstruction can cause an attenuation of up to \SI{15}{\decibel} in practical deployments, where several meters or more separate the \gls{hmd}, obstructing person, and \gls{ap} \cite{gustafson_characterization_2012}. Similarly, the shadow fading parameter of statistical models for indoor deployments, such as those part of 3GPP~38.901, suggests that attenuation due to shadowing is typically below \SI{10}{\decibel}, and that it can potentially reach up to \SI{15}{\decibel}.
\arxiv{Based on the measurement data, we can assess external obstruction in a potential multi\nobreakdash-user scenario. Consider there are several playspaces located next to each other, each with its respective user, and that one or more \glspl{ap} are mounted to the surrounding walls. The highest link obstruction probability is obtained when the centres of two playspaces and an \gls{ap} align. To determine how often \gls{los} obstruction occurs in such a deployment scenario, we first approximate a person's head and upper body using a vertical cylinder with a radius of \SI{15}{\centi\meter} and \SI{40}{\centi\meter}, respectively.
Next, we assess the distance of users from their mean position in the playspace. 
In particular, we consider \piwi since users exhibit the least amount of lateral movement, resulting in the highest degree of shadowing when the two playspaces and the \gls{ap} are aligned. In \piwi, users spend \SI{7}{\percent} of time within a \SI{7.5}{\centi\meter} distance from the central position and \SI{32}{\percent} of time within a distance of \SI{20}{\centi\meter} (the corresponding percentiles, P$_{7}$ and P$_{32}$, are not shown in \Cref{tab:mobility_characterisation}, ID A). Hence, the two numbers respectively represent the worst\nobreakdash-case external obstruction probability when link shadowing occurs due to either the head or torso. Determining which of the two applies in a particular scenario further requires information about user body height, user stance, and \gls{ap} mounting height. The shadowing probability increases if more users are considered.}

\emph{Self\nobreakdash-shadowing} happens when the \gls{hmd} user obstructs the \gls{los} with one or both hands. Although the attenuation caused by a hand or an arm is in general less severe than the attenuation upon obstruction by a person's torso, according to the \gls{gtd}, it could exceed \SI{5}{\decibel} due to the hand's vicinity to the \glsposs{hmd} antennas.
\arxiv{According to the measured mobility, self\nobreakdash-obstruction, due to raising one's hands above the \gls{hmd}, has a probability of approximately \SI{3}{\percent}, making it far less likely than the above\nobreakdash-described external shadowing. However,  the user's hands can remain above the head for up to \SI{45}{\second}, possibly causing long\nobreakdash-term link obstruction. Moreover, the relative proximity of the hands to the \gls{hmd}, reported in \Cref{sec:planar_hand_position}, means that a raised hand covers a noteworthy amount of the \glsposs{hmd} angular view. For example, a \SI{7}{\centi\meter} wide palm would obstruct \SI{20}{\degree} of the \glsposs{hmd} view if placed at the median measured distance from the \gls{hmd}, i.e., at \SIrange{20}{30}{\centi\meter}.}

Providing alternative propagation paths between the \gls{hmd} and \gls{ap} can circumvent external~and self\nobreakdash-shadowing. Therefore, multiple antennas should be placed at different positions on the \gls{hmd}, while several \gls{ap} antennas should be distributed in the environment if high link reliability is required. Conversely, consumer deployments might often feature a single \gls{ap} due to financial reasons, while \gls{ap} deployment height is usually limited by a low ceiling (e.g., residential buildings). Moreover, mixing 
users of various body heights in such deployments is likely to lead to some degree of shadowing.

\subsection{Antenna and beam misalignment}
\label{sec:impact_on_mmwave:misalignment}

\Gls{mmw} propagation undergoes substantial path loss and is subject to pronounced attenuation during shadowing. To increase gain (mitigate loss), directional antennas with a narrow radiation pattern are employed. Moreover, multiple antennas are assembled into arrays, which yields an even narrower instantaneous radiation pattern, referred to as a beam. Since the antenna arrays are fixed onto the \gls{hmd}, which moves relative to a stationary \gls{ap}, the mobility of the user can lead to so\nobreakdash-called \emph{antenna} and \emph{beam misalignment}.

\emph{Antenna misalignment} 
occurs when the high\nobreakdash-gain regions of the antenna radiation pattern are rotated away from the \gls{los} component. 
For example, commonly\nobreakdash-employed microstrip patch antennas have a directivity gain of several \SI{}{\decibel}, which comes at the price of poor reception from angles exceeding $\pm\mathrm{90}^\circ$ from boresight.
Clearly, the \gls{hmd} requires more than one such antenna to cover the entire horizon (azimuth). However, even an \gls{hmd} with several antennas, distributed along the azimuth, would observe a varying gain during rotation due to the uneven angular radiation pattern. Furthermore, the large pitch bias observed in this measurement campaign and in prior art suggests that antenna misalignment is likely to also occur in the elevation domain. Certain orientation correction factors, applied during antenna placement on the \gls{hmd}, could mitigate the impact of this bias on antenna misalignment.

\emph{Beam misalignment} is the result of antenna array rotation, which is not counteracted using beam steering. \Gls{mmw} Wi\nobreakdash-Fi, that is, the IEEE~802.11ad/ay standard, typically conducts beam steering once per beacon interval, i.e., approximately ten times per second. An \gls{hmd} employing beams with a \SI{10}{\degree} half\nobreakdash-power beam width and rotating at \SI{100}{\degree\per\second} would observe a gain difference of at least \SI{3}{\decibel} during a \SI{0.1}{\second} beacon interval, depending on how well the beam was aligned initially. To cope with the fast rotation, the link layer on \gls{mmw} \glspl{hmd} could adapt the beam steering behavior based on user mobility (e.g., using \gls{imu} data), which often follows predeterministic patterns. For example, the \gls{hmd} could conduct additional beam steering during periods of elevated mobility or it could proactively steer (or switch) beams based on short\nobreakdash-term mobility inference, without occupying airtime with beam probing.

Losses due to antenna and beam misalignment can further reduce the gain of a \gls{mmw} link, in addition to the attenuation caused by shadowing. The expected gain reduction of a system with sufficient antennas and proactive beam steering is in the range of several \SI{}{\decibel}, which can increase if insufficient antennas are employed and if inadequate beam steering is provided. Both losses heavily depend on the employed hardware and the mobility characteristics of the~\gls{xr}~use~case at hand.

\section{Conclusion and outlook}
\label{sec:conclusion}

In this paper, we have presented an open \gls{6dof} \gls{xr} tracking dataset (available at \cite{cacerumd_zenodo}) and the mobility characterisation results, listed in \Cref{tab:mobility_characterisation}. The tracking data were sampled from two \gls{xr} games and one \gls{xr} training application, all of which feature possibly adverse mobility profiles, hence, making the dataset relevant for studies concerning wireless \glspl{hmd}. The characterisation also proposes statistical distributions that best fit the measured mobility quantities. We have found that user head rotation can surpass \SI{360}{\degree\per\second} in all three applications and that users exhibited a considerable head pitch bias, ranging up to \SI{32}{\degree} (looking down). 
Moreover, the study revealed that, while the probability of users raising their hands above the head is less than \SI{3}{\percent}, they can sustain this posture for durations of up to \SI{45}{\second}. The raised hands severely deteriorate \gls{mmw} link quality, although, hands pose a lesser adversity than in mobile telephony. Instead, \gls{xr} use cases are characterised by the rapid rotation.
Our assessment of the possibly adverse effects of the recorded mobility on \gls{mmw} links, summarised in \Cref{tab:mobility_effects_on_mmwave}, has shown that mobility poses a substantial challenge in deploying \gls{mmw} networks for \gls{xr}. Namely, due to external shadowing caused by third persons, self\nobreakdash-shadowing with one's hands, misalignment of the antenna radiation patterns, and misalignment of array beams.

We began the measurement campaign using a \gls{mmw} \SI{60}{\giga\hertz} wireless adaptor on the \gls{hmd} and switched to a tethered connection after reports of user discomfort during gameplay, indicating that the current \gls{mmw} \glspl{hmd} are not providing the desired user experience. As highlighted, part of the problem may lie in inadequate beam steering to cope with user mobility. The generated database can be used as\nobreakdash-is to evaluate novel link management algorithms, such as proactive beam steering (or selection) based on mobility inference.
Furthermore, the statistical approximation of the evaluated mobility parameters can be extended to build a mobility model for high\nobreakdash-mobility applications, potentially including all \gls{6dof} and user hands in addition to the head. Lastly, the \gls{ssq} and tracking measurements could together shed additional light on the connection between user sickness and mobility.

\section*{Acknowledgement}
{\footnotesize%
\noindent
\footnotesize{Thanks to our volunteers for making the experiment possible.}
\vspace{0.5 mm}
 
\noindent
\begin{minipage}{0.2\linewidth}
    \vspace{1 mm}
    \includegraphics[width=0.9\columnwidth]{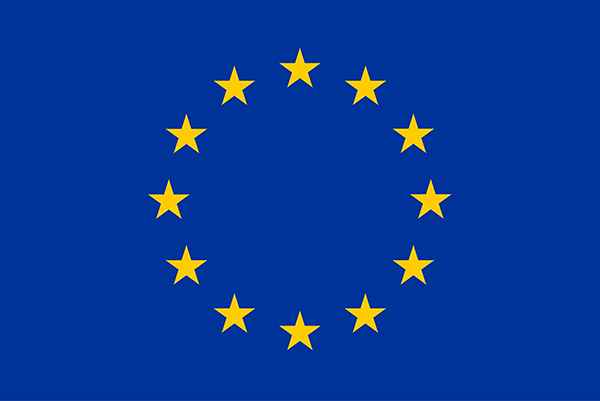}
\end{minipage} \hfill
\begin{minipage}{0.8\linewidth}
    \vspace{2 mm}
    \footnotesize{This work has received funding from the EU’s Horizon 2020 and Horizon Europe programmes under grant agreements No.~861222~(MINTS) and 101096302~(6GTandem).}
\end{minipage}
}

\printbibliography

\narxiv{
\begin{IEEEbiographynophoto}{Alexander Marin\v{s}ek}
is a doctoral student at KU Leuven, Belgium, working on \gls{mmw} \glspl{hmd}. His expertise and interests lie within the wireless technology, extended reality, and embedded systems domains.
\end{IEEEbiographynophoto}

\begin{IEEEbiographynophoto}{Sam De Kunst}
is a recently-graduated ICT student from KU Leuven and employee of Skeyes, Belgium. His expertise and interests are within the virtual reality, software networking and VoIP domains.
\end{IEEEbiographynophoto}

\begin{IEEEbiographynophoto}{Gilles Callebaut}
is a post-doctoral fellow at KU Leuven, who focuses on minimizing IoT device energy consumption. His goal is to shift communication paradigms towards sustainable, low-power systems.
\end{IEEEbiographynophoto}

\begin{IEEEbiographynophoto}{Lieven De Strycker}
is a KU Leuven professor. He founded the DRAMCO research group in 2001 and has since coordinated it in various international projects, covering topics from acoustics to wireless technology and beyond.
\end{IEEEbiographynophoto}

\begin{IEEEbiographynophoto}{Liesbet Van der Perre}
is a full professor at KU Leuven and guest professor at Lund University, Sweden. She leads the DRAMCO laboratory in several well\nobreakdash-renowned international projects, mainly focusing on wireless technology.
\end{IEEEbiographynophoto}

}

\end{document}